\documentclass[12pt]{iopart}

\usepackage{graphicx}                       
\usepackage{geometry} 
\usepackage{amssymb, latexsym}

\usepackage{iopams}  
\begin{document}

\title[PIV measurements using refraction at a solid/fluid interface.]{PIV measurements using refraction at a solid/fluid interface.}

\author{Damien Cabut$^{1}$, Marc  Michard$^{1}$, Serge Simoens$^{1}$, Violaine Todoroff$^{2}$, Corentin Hermange$^{2}$ and Yohan Le-Chenadec$^{2}$}

\address{$^{1}$ Univ Lyon, Ecole Centrale de Lyon, INSA Lyon, Université Claude Bernard Lyon I, CNRS, LMFA, UMR 5509, 36 Avenue Guy de Collongue, F-69134, ECULLY, France.\\
$^{2}$ Manufacture Fran\c{c}aise des Pneumatiques Michelin, Clermont-Ferrand, France}
\ead{damien.cabut@ec-lyon.fr}
\vspace{10pt}
\begin{indented}
\item[]July 2020
\end{indented}

\begin{abstract}
Laser light sheet refraction at the fluid flow/window interface is proposed to proceed fluorescent PIV measurements, in particular for cases for which optical access limits the possibilities of illuminating a flow with a planar collimated light sheet. The study of the laser sheet propagation for the present measurement method is led with ray tracing to quantify the illumination intensity. To describe the limitations of this developed PIV version (the R-PIV), we proposed a cross-correlation model (CCM) for analysing image pairs and that is an extension of the ones used in the cases of $\mu$-PIV or P-PIV techniques. This cross-correlation model accounts for the R-PIV optical constraints as 1) in-depth and longitudinal inhomogeneities of the light sheet refraction at the location of the object plane or 2) integration in depth of the fluorescence from seeding particles. This refracted PIV (R-PIV) and the P-PIV techniques are further compared in the case of an academic pipe flow whose analytic solution is known. Our extended cross-correlation model function is finally applied, knowing the theoretical flow velocity, the flow illumination and the seeding flow parameters, to analyse some error levels in determining the real flow velocity field.
\end{abstract}

%
\vspace{2pc}
\noindent{\it Keywords}: Particle Image Velocimetry, Refraction, Fluid flow
%
%
%
%

\section{Introduction.}

The particle image velocimetry (PIV) is a well known technique used in fluid mechanics to measure velocity fields in fluid flows (Adrian and Yao 1985 \cite{adrian1985pulsed}, Adrian 1991 \cite{adrian1991particle}). In PIV, particles are seeding the fluid flow of interest and are illuminated by a double pulsed laser light sheet. Images of these particles are recorded on a camera at two successive instants, $t$ and $t+dt$ (corresponding to the pulse generation times), and the velocity field is calculated from correlation techniques, as cross-correlation, between two successive recorded images.

\subsection{Planar PIV}

For centimetric field of view with planar PIV measurements (denoted as P-PIV, in the following), the working distance is usually a macroscopic scale. With these conditions, the depth of focus of the lens mounted on the camera is proportional to the working distance (Kingslake 1992 \cite{kingslake1992optics}) and large (at least few millimeters) compared to the light sheet thickness. Therefore, the depth over which particles have an image that contributes to the cross-correlation is driven by the thickness of the laser sheet illuminating the field of view.

For P-PIV measurements in a channel flow, the laser sheet should be as thin as possible and ideally perfectly parallel to the camera recording plane to avoid distortion of the images. A constraint to use this technique is that the laser sheet should be perpendicular to the camera axis and thus have to penetrate inside the field of view perpendicularly to the wall supporting the flow of interest, and from which the flow view is done. 

If these prerequisites are respected, two successive laser pulses illuminate seeding particles with a time delay $dt$ and two images are recorded. The displacement $\textbf{dx}$ of particles (hypothetically homogeneous) between both images can be calculated using a cross-correlation algorithm. Therefore the velocity field is determined with $\textbf{U}=\frac{\textbf{dx}}{dt}$. 

After recording an image, a digital matrix of pixels is obtained whose grey-levels describe the intensity re-emitted by the particles from the field of view. For two successive images, the digital cross-correlation is applied to small parts of the images called interrogation windows with a size ($\delta w_{x}, \delta w_{y}$) in pixels (($\Delta w_{x}, \Delta w_{y}$) in mm) depending on the displacement of particles. These windows must be small enough to consider that all the particles within the window have the same velocity. The cross-correlation algorithm consists in finding the pattern produced by the particle images in an interrogation window at time $t$, inside the same interrogation window at the time $t+dt$ (Adrian and Yao 1985 \cite{adrian1985pulsed} and Adrian 1991 \cite{adrian1991particle}).

If we consider $I_{1}(X_{W}+x,Y_{W}+y)$ ($(x,y)\in [X_{W}-\delta w_{x}/2,X_{W}+\delta w_{x}/2]* [Y_{W}-\delta w_{y}/2,Y_{W}+\delta w_{y}/2]$), as describing the intensity pattern in the interrogation window centred in $(X_{W},Y_{W})$ in image 1 and $I_{2}(X_{W}+x,Y_{W}+y)$ as describing the intensity pattern in the interrogation window centred in $(X_{W},Y_{W})$ in image 2, the cross-correlation between $I_{1}$ and $I_{2}$ at $(X_{W}+\delta x,Y_{W}+\delta y)$ is calculated according to Scarano and Riethmuller 2000 \cite{scarano2000advances} as:

\begin{eqnarray}
R_{12}(\delta x,\delta y)&=&\int_{X_{W}-\Delta w_{X}/2}^{X_{W}+\Delta w_{X}/2}\int_{Y_{W}-\Delta w_{Y}/2}^{Y_{W}+\Delta w_{Y}/2}I_{1}(X_{W}+x, Y_{W}+y)\ldots \nonumber \\
&\cdot & I_{2}(X_{W}+x+\delta x,Y_{W}+y+\delta y) dxdy
\label{eq21}
\end{eqnarray}

For P-PIV applications and in the present work, the cross-correlation uses a discrete formulation following the pixel intensity distribution over the used camera sensors. The discrete cross-correlation function is calculated as expressed by Willert and Gharib 1991 \cite{willert1991digital} as :

\begin{eqnarray}
R_{12}(\delta  x_{px},\delta  y_{px})=\sum_{k=-\delta w_{x}}^{\delta w_{x}}\sum_{n=-\delta w_{y}}^{\delta w_{y}} I_{1}(k, n)\cdot I_{2}(k+\delta x_{px}, n+\delta y_{px})
\end{eqnarray}
where $\delta x_{px}$ and $\delta y_{px}$ are displacement variables between image 1 and 2 in pixels and if we consider an interrogation window centred in ($0$, $0$) in image 1.

\begin{figure}
    \centering
    \includegraphics[width=0.55\textwidth]{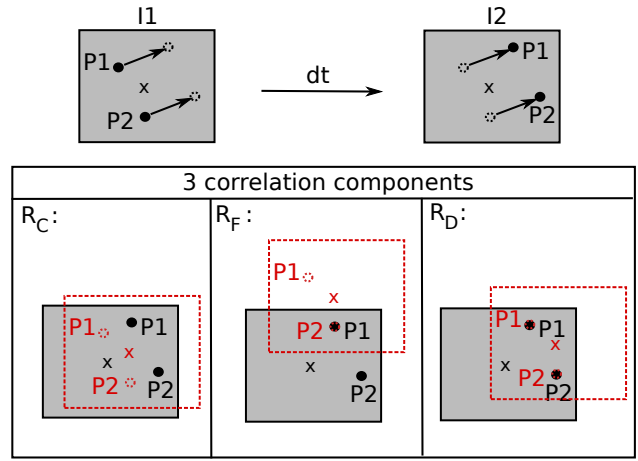}
    \caption{Scheme of the decomposition of the cross-correlation function.}
    \label{fig1}
\end{figure}

To determine the particle displacement in such an interrogation window, the displacement ($\delta x_{px}$, $\delta y_{px}$) maximising the cross-correlation function is found. According to Kean and Adrian 1992 \cite{keane1992theory}, the cross-correlation function can be decomposed into 3 parts. $R_{C}$ is between backgrounds of image 1 and image 2. $R_{F}$ takes into account the correlation of a particle in image 1 with an other particle in image 2. $R_{D}$ is the displacement component which corresponds to the correlation of particles in the image 1 with their own images in image 2. These components are sketched in Fig.\ref{fig1}.

This leads to a cross-correlation function between image 1 and 2 that is the sum of these 3 components (Adrian 1988 \cite{adrian1988double}).

\begin{eqnarray}
R_{12}={R_{C}}_{12}+{R_{F}}_{12}+{R_{D}}_{12}
\label{eq23}
\end{eqnarray}

\begin{figure}
    \centering
    \includegraphics[width=1\textwidth]{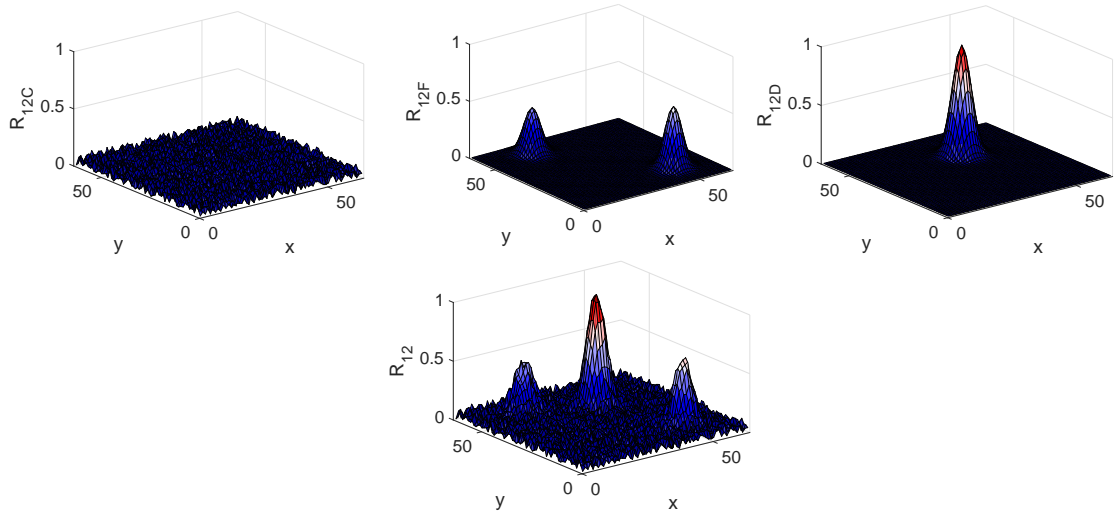}
    \caption{Illustration of the 3 components contributions to the total cross-correlation.}
    \label{fig2}
\end{figure}

An example of the cross-correlation and its components is displayed in Fig.\ref{fig2}.

In the example of Fig.\ref{fig1} and Fig.\ref{fig2}, the two particles have the same displacement. They both contribute with the same amplitude to ${R_{D}}_{12}$. On the contrary, the correlation between particle 1 and 2 contributes symmetrically to two different peaks in ${R_{F}}_{12}$. With two particles, the ratio between ${R_{D}}_{12}$ and ${R_{F}}_{12}$ peaks is $2$. If we consider an higher number of particles in the interrogation window in image 1, the ratio between ${R_{D}}_{12}$ and ${R_{F}}_{12}$ grows proportionally.

However, P-PIV conditions are difficult to respect in particular cases for which the flow container does not possess two optical accesses or if one is too far from the field of interest (thin fluid film moving on a flat plate, tire rolling on a water puddle, etc).

\subsection{Microscopic PIV}

For $\mu$-PIV, the characteristic scales of the studied flow is of a sub-millimeter range. Therefore, it is difficult to obtain light sheets thin enough compared to the flow scales. For such a configuration, microscopic lenses with large magnifications are used. Working distance and depth of focus are consequently very small. In $\mu$-PIV the selection of the measured position in the flow is then mainly driven by the collection optic properties instead of the laser sheet thickness (Vetrano et al. 2008 \cite{vetrano2008applications}).

However in this different optical context, the acquisition process remains the same than for the P-PIV. Two laser pulses separated by a time step $dt$ illuminate the seeding particles and two images are recorded.

In confined configurations as for $\mu$-PIV, with low velocities, fluorescent particles are often used (as used by Brücker 2000 \cite{brcker2000piv}). The fluid illumination is global, interfering reflections introduce background noise that is reduced by the use of fluorescence. The drawback is that the intensity of the light reemitted is smaller which makes it more complex to use in high speed cases for which recording time (and thus recording intensity duration) is decreasing.
However in this case the depth over which particles appear on the images is driven by the recording optics. The particles appearing on the images are only particles at the specific location of the object plane and the cross-correlation properties are the same as used for the P-PIV.

Some studies focus on the effect of a thick depth of focus for $\mu$-PIV configuration (Olsen and Adrian 2000 \cite{olsen2000out}, Meinhart et al. 2000 \cite{meinhart2000volume}). These studies take into account the integration effect due to the fact that particles at different heights appear on the images. The cross-correlation takes into consideration the contribution of the particles in the whole illuminated volume depending on their focusing (as studied by Kloosterman et al. 2011 \cite{kloosterman2011flow}, Fouras et al. 2009 \cite{fouras2009volumetric}, Willert et al. 1992 \cite{willert1992threes} and Pereira et al. 2000 \cite{pereira2000defocusing}). All particles are here illuminated with the same incident intensity.

\subsection{The refracted PIV as an hybrid technique.}

The main purpose of the present work is to measure the velocity field in a puddle with a rolling tire. The field of view targeted is around $200$ mm large with a minimum working distance of $500$ mm. Moreover, the macroscopic scale of the targeted flow does not allow the application of $\mu$-PIV technique. For this measurement configuration, the laser head can not be placed at the ground level for safety considerations. It is not possible to perform P-PIV for this type of semi-confined flow (free surface). Therefore we develop a method in order to perform PIV measurements fo such macroscopic scale flows with only one optical access for both emitting and receiving optics.

\begin{figure}
    \centering
    \includegraphics[width=0.7\textwidth]{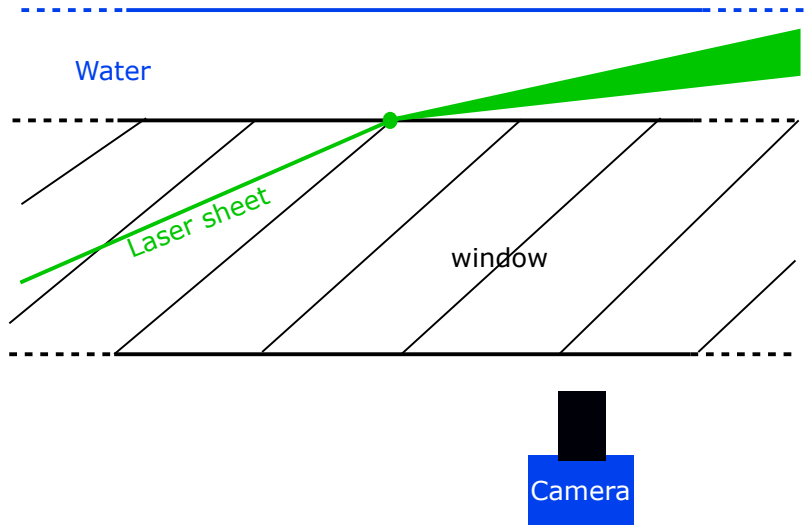}
    \caption{Scheme of the illumination method for velocity measurements inside a liquid film using refraction.}
    \label{fig3}
\end{figure}

The solution chosen in the present work to perform PIV measurements with one optical access at macroscopic scales is based on the refraction of the laser beam at the window/flow interface allowing illumination of the seeding particles (Fig.\ref{fig3}). This extension of the PIV with a refraction of the light sheet is here called R-PIV. Refraction has previously been used by Minami et al. 2012 \cite{minami2012observation} to study a water droplet behaviour in a rubber/window contact patch area with a global illumination.

In this paper, an academic flow is studied in order to identify and quantify the potential biases induced by this technique. The academic flow has been chosen to allow 1) comparison with an analytical solution in laminar regime and 2) comparison with standard P-PIV measurements.

The experimental set-up for the validation of the method is presented in Sec.\ref{ES} with measurements of the optic characteristics used for the emission and the reception. In Sec.\ref{OP}, specific optical properties of the R-PIV technique are discussed. Thereafter, a Cross-Correlation model (CCM) between particle images including a non homogeneous illumination is presented in Sec.\ref{CC}. This model supports a discussion on the theoretical precision of planar PIV (P-PIV) technique classically used for macroscopic transparent containment flows. Finally R-PIV measurements are compared to a reference flow field and to results obtained when applying CCM to this flow configuration Sec.\ref{MR}.

\section{Experimental set-up.}
\label{ES}

\subsection{The bench and operating conditions}

\begin{figure}
    \centering
    \includegraphics[width=0.7\textwidth]{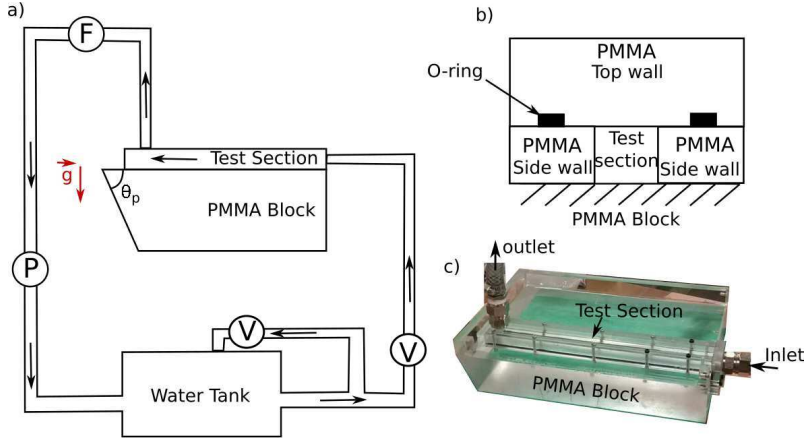}
    \caption{a) Scheme of the hydraulic loop with P for the pump, V for valves and F for the flow-meter. b) Scheme of a cross-section of the channel. c) Picture of the channel mounted on the PMMA block.}
    \label{fig4}
\end{figure}

In order to characterize this extension of PIV, a channel flow with water as the working fluid is used. The flow is developing in a square channel inside an hydraulic loop depicted in Fig.\ref{fig4}. This loop is composed of :
\begin{itemize}
\item a water-tank,
\item a pump to set the fluid in movement,
\item a flowmeter to measure the flow rate,
\item a transparent test section,
\item a by-pass system to regulate the flow rate, 
\end{itemize}

The test channel is composed of PMMA transparent top and side walls. The channel is mounted on a PMMA block with a prismatic shape (Fig.\ref{fig4} c)). The upper part of this prism constitutes also the bottom part of the channel. The height of the square cross-section of the channel is $h=8$ mm. The thickness of the prismatic PMMA block is $49$ mm. This block has an inclined face near the channel exit with an inclination of $\theta_{p}=64^{\circ}$ (Fig.\ref{fig4}). This angle is chosen to be close to the critical angle given by Snell-Descartes Laws for a total reflection at the interface PMMA/water with $n^{PMMA}=1.49$ and $n^{water}=1.33$.

\begin{figure}
    \centering
    \includegraphics[width=0.9\textwidth]{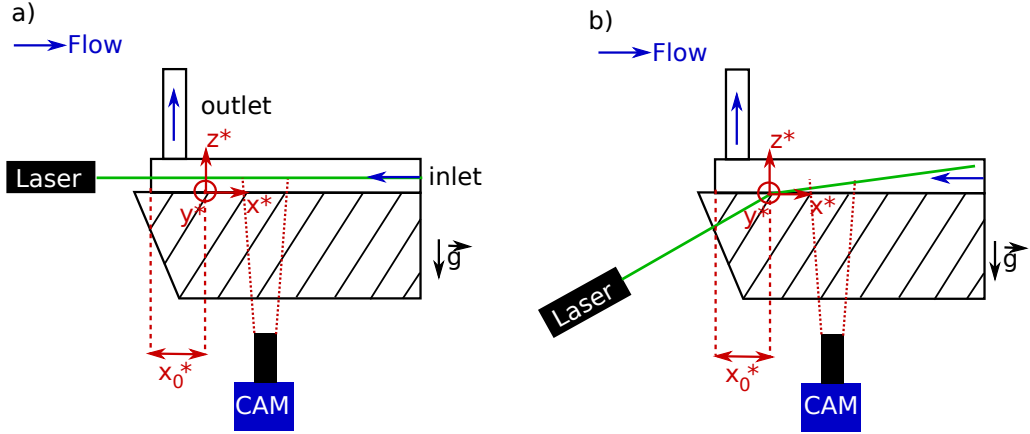}
    \caption{Side view of the experimental setup for a) P-PIV, b) R-PIV. In all the following study, $x^{*}=\frac{x}{h}$, $y^{*}=\frac{y}{h}$, $z^{*}=\frac{z}{h}$.}
    \label{fig5}
\end{figure}

The outlet of the test section is perpendicular to the main flow direction in order to allow illumination through the end wall of the channel (as sketched Fig.\ref{fig5} a)). The upper plate of the channel is removable to perform specific optical arrangements inside the channel when the fluid is at rest, as to dispose calibration targets. All the coordinates are normalized in this work by the height of the channel $h$ (respectively $x^{*}=\frac{x}{h}$, $y^{*}=\frac{y}{h}$ and $z^{*}=\frac{z}{h}$).

The origin of the coordinate system is considered at a distance $x_{0}^{*}=7$ upstream the end of the channel. This distance corresponds to the distance between the end of the channel and the emerging point of the laser beam in the water for the refracted configuration (Fig\ref{fig5}b)). The field of view is placed at a distance $L_{c}^{*}=18.5$ downstream of the channel inlet.

Measurements described here are performed for a Reynolds number based on the bulk velocity, $U_{b}$, deduced from the flow rate, and the channel height, $h$, equal to $465$. This value is small enough for the flow to be laminar. An analytical solution of the streamwise velocity component $U$ ($V=0$ and $W=0$) for a fully developed steady laminar flow in a square channel is given, in F.M. White and Corfield 2006 \cite{white2006viscous}. We use here a non dimentionalised form of this solution $U^{*}=U/U_{b}$, where $U_{b}$ is the bulk velocity calculated from the Reynolds number, the height of the channel is the spatial characteristic and $(1/2 \rho U_{b}^2 )$ is the pressure characteristic :

\begin{eqnarray}
U^{*}(y^{*},z^{*})=\frac{2\rho hU_{b}}{\mu \pi^{3}}\left(\frac{-d\hat{p}^{*}}{dx}\right)\cdot\nonumber \\
\sum_{i=1,3,5...}^{\infty}\left((-1)^{(i-1)/2}\left[1-\frac{cosh(i\pi (z^{*}-0.5))}{cosh(i\pi /2)}\right]\frac{cos(i\pi (y^{*}-0.5))}{i^{3}}\right)
\label{White}
\end{eqnarray}
where $V^{*}=0$ and $W^{*}=0$, $\mu$ is the dynamic viscosity and cosh(.) and tanh(.) respectively the cosinus and tangent hyperbolic functions. The pressure is normalised as $\hat{p}^{*}=\hat{p}/((1/2)*(\rho U_{b}^2))$ and its axial gradient is related to $U_{b}$, $h$, $\rho$ and $\mu$ by :

\begin{eqnarray}
\frac{d\hat{p}^{*}}{dx^{*}}=\frac{24\mu}{\rho U_{b} h}\frac{1}{\left[1-\frac{192}{\pi^{5}}\sum_{i=1,3,5...}^{\inf}\frac{tanh(i\pi /2)}{i^{5}}\right]}\nonumber
\end{eqnarray}

\subsection{Emitting Optics}
\label{EO}

The test bench is designed in order to allow the use of both P-PIV (Fig\ref{fig5}a)) and R-PIV (Fig.\ref{fig5}b)). The P-PIV is performed with a laser sheet parallel to the flow direction and entering the channel by the end wall of the test section. The camera in this configuration is placed perpendicular to the light sheet and below the prismatic shaped block. In the refracted configuration (Fig.\ref{fig5}b)), the incident light sheet is inclined with a small angular offset from the normal of the inclined face of the PMMA block. After the refraction at the flow/window interface, the purpose is to obtain the slightly inclined laser beam from the horizontal (around $5^{\circ}$ in this paper).

The laser source used is a double pulsed Laser:Yag (Litron Bernouilli) emitting at a wave length of $532$ nm. The output diameter of the laser beam is $6$ mm with a divergence angle of about $3.5$ mrad. The laser sheet is generated with a specific optical device composed of a beam-expander, with an expansion factor of $2$, a spherical converging lens with $f=500$ mm and a cylindrical diverging lens with $f=400$ mm to spread the laser sheet in the $y^{*}$ direction. For the planar configuration (Fig\ref{fig5}a)), the spherical lens allows us to focus the laser beam in the measurement area to minimise the thickness of the laser sheet.

For both optical configurations, the laser head is placed on translation stages to set the axial and vertical ($x^{*}$ and $z^{*}$) positions of the laser head with a precision of $0.01$ mm. The emitting optics is also placed on a rotation stage with a precision of $0.01^{\circ}$ in order to finely tune the inclination of the laser sheet for the refracted configuration (Fig.\ref{fig5}b)).

The distance used here between the emission optic and the inclined face of the prism is approximately $270$ mm. The total thickness of the prism being $49$ mm, the impingement of the leaser sheet with the water/PMMA interface is in the converging propagation part of the beam (before the beam-waist).

\subsection{Seeding}
\label{RO}

With this refracted illumination technique, spurious background light is generated by multiple reflections and refractions at the different PMMA/air and PMMA/water interfaces. To remove this spurious background light, flow seeding is achieved using fluorescent particles homogeneously mixed to the water in the water tank. The particles used here are Lefranc et Bourgeois paint as characterized by Strubel et al. 2017 \cite{strubel2017fluorescence} and Nogueira et al. 2003 \cite{nogueira2003simultaneous}, with a particle diameter around $d_{p}=10 \mu m \pm 4$ $\mu$m. The relaxation time of these particles due to viscous drag, is $\tau_{v}=\frac{\rho_{p} d_{p}^{2}}{18\rho\nu}=6.10^{-6}$ s. For the laminar flow in the channel studied here, the characteristic time of the flow is defined as the convection time from the channel inlet to the measurement area (this corresponds to the convection time from inlet disturbances to reach the field of view): $\tau_{0}=\frac{L_{c}}{U_{b}}=2.55$ s (with $L_{c}=148$ mm, distance between the measurement area and the channel inlet). Therefore the Stokes number of these particles in the channel flow is $St=\frac{\tau_{v}}{\tau_{0}}=2.35\cdot 10^{-6}$. This small value ensures a good agreement between the particle velocities and the fluid flow velocity.

When these fluorescent particles are illuminated by a $532$ nm light sheet, the fluorescence spectrum is centred around $570$ nm. The lens of the camera is equipped with an optical band-pass filter centred at $586$ nm with a width of $\pm 20$ nm to record the fluorescing particle images free from reflected noise at the laser wavelength ($532$ nm).

\subsection{Image acquisition and processing}

Image acquisitions are made with a double frame sCMOS camera. The camera sensor, of size $14$x$16.6$ mm, is composed of $2160$x$2560$ pixels. The synchronisation of the laser pulses and the camera acquisitions is achieved with a synchronisation unit (PTU). The acquisition is made with a commercial software (Lavision, Davis 8). The maximal acquisition rate is limited by the laser frequency at $15$ Hz.

For these measurements, a NIKKOR-NIKON lens of $f=135$ mm focal length is used with two extension rings of length $12$ and $32$ mm, with a working distance of $500$ mm. The dimensions of the field of view are $\Delta_{x}^{*}=5.28$ and $\Delta_{y}^{*}=4.46$ for a magnification factor of $M=0.39$. This measurement area is placed $148$ mm downstream the channel inlet as mentioned previously and $80$ mm upstream the channel outlet.

In order to improve the signal/noise ratio $500$ independent image pairs are recorded. In this paper, the individual cross-correlation functions are averaged instead of the instantaneous velocities. The size of the interrogation windows used here is decreasing from $64$x$64$ pixels down to $32$x$32$ pixels. Therefore the final size of the interrogation windows normalized by the channel height is ${d_{w}}_{x}^{*}={d_{w}}_{y}^{*}=0.066$.

\section{Optical properties of R-PIV}
\label{OP}

\subsection{The laser sheet propagation.}

\subsubsection{Measurement of the laser sheet intensity after the refraction.}
\hfill \break
\label{Profmes}

\begin{figure}
    \centering
    \includegraphics[width=0.9\textwidth]{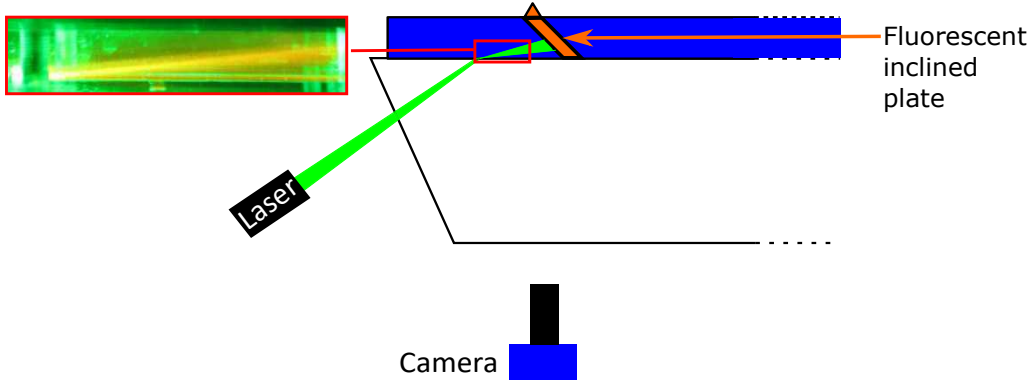}
    \caption{Set-up for the measurement of the inclined light sheet intensity profile. Left picture is a photo of the emerging light sheet.}
    \label{fig6}
\end{figure}

With the present specific R-PIV technique, the shape of the laser sheet after refraction is a priori unknown. An intensity profile measurement technique is developed in order to quantify the thickness of the laser sheet and its structure along the propagation direction in the channel. The measurements are done when the fluid is at rest by positioning in the light sheet path a fluorescent target with a flat face inclined at $45^{\circ}$ with the horizontal lower wall (Fig\ref{fig6}). Fluorescent light emitted by the impingement of the laser sheet on the plate is collected by the camera placed below the PMMA block (Fig\ref{fig6}). The intensity is spatially averaged along the spanwise $y^{*}$ direction and studied as a function of the $x^{*}$ direction along the plate, to deduce the intensity profile $I_{0}^{*}(z^{*})$ due to the inclination of the plate. To know the evolution of the intensity profile in the channel, the measurement is repeated at different locations $x^{*}$ of the plate.

\begin{figure}
    \centering
    \includegraphics[width=1\textwidth]{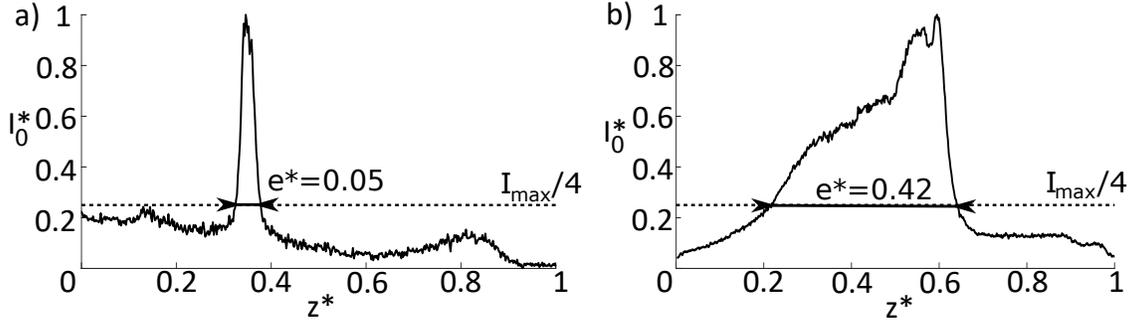}
    \caption{Light sheet profile $I_{0}^{*}(z^{*})$ a) near the beam-waist for P-PIV, b) at $x^{*}=8$ for the R-PIV.}
    \label{fig7}
\end{figure}

The light sheet intensity profile measurement is performed for both optical configurations (Fig.\ref{fig5} a) and b)). A typical normalized intensity profile $I_{0}^{*}(z^{*})$ is presented in Fig.\ref{fig7} for both optical configurations (P-PIV and R-PIV).

For the P-PIV configuration, the intensity measurements are made at $x^{*}=6$ (near the beam-waist position for Fig.\ref{fig5} a)). As expected, the intensity profile appears to be sharp and nearly symmetric around its peak (Fig.\ref{fig7}a)). The light sheet thickness $e$ is arbitrarily defined as the profile width at height $I_{0}^{*}=0.25$. Its value is here $e=0.4$ mm which corresponds to $e^{*}=0.05$ in its normalised form which is small enough to consider the laser sheet thickness negligible facing the height of the channel.

For the R-PIV technique, the measurement presented in Fig.\ref{fig7}b) is made at $x^{*}=6.5$. This profile shows two main features, a much larger thickness and a strong asymmetry. Using the same definition as for the thickness of the peak for the P-PIV profile, its value is $e^{*}=0.42$. Therefore, we expect the large thickness of the laser sheet to introduce greater bias than for the P-PIV arrangement.

\begin{figure}
    \centering
    \includegraphics[width=0.8\textwidth]{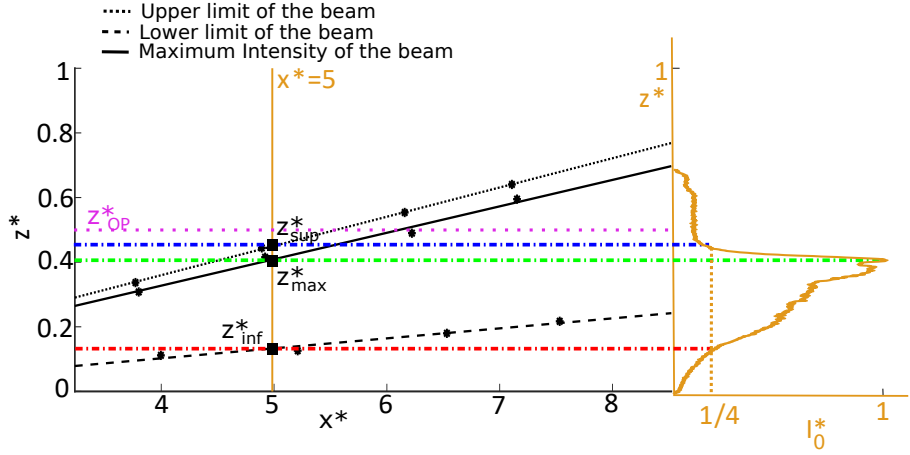}
    \caption{Dashed-doted colored lines are planar horizontal light sheets corresponding to $z^{*}_{sup}$,$z^{*}_{max}$ and $z^{*}_{inf}$. Purple dotted line is the position of the camera object plane. The normalized intensity profile $I(z^{*})$ (right) is also presented to show the correspondence between limit lines and the intensity distribution, in the case $x^{*}=5$.}
    \label{fig8}
\end{figure}

For the R-PIV technique, the measurement of the intensity profile is repeated at four different locations $x^{*}$ along the measurement area. This allows to study the evolution of the laser sheet edges, ${z_{inf}}^{*}$ and ${z_{sup}}^{*}$ (defined as the value of $z^{*}$ where $I_{0}^{*}=0.25$) and the maximum intensity of the laser sheet ${z_{max}}^{*}$ (Fig.\ref{fig8}).

The laser sheet appears to evolve with a small inclination in the channel (approximately $5^{\circ}$) with a growing thickness along the measurement area $e^{*}(x^{*})$. The normalized intensity profile $I_{0}^{*}(z^{+})$ (with $z^{+}=\frac{z^{*}}{e^{*}(x^{*})}$) is self-similar at every position $x^{*}$.

\subsubsection{Ray tracing model.}
\hfill \break
\label{Ray}

A ray tracing model is used here to explain the results obtained for the measurements of the laser sheet thickness. The propagation problem is here simplified as a 2D problem in the ($x^{*}$, $z^{*}$) plane. In this ray model, the initial cross-section shape of the laser beam is modeled as a set of $100000$ rays equally spaced. The intensity distribution is set with respect to have a complete beam with a Gaussian intensity profile of width $19$ mm.

\begin{figure}
    \centering
    \includegraphics[width=0.7\textwidth]{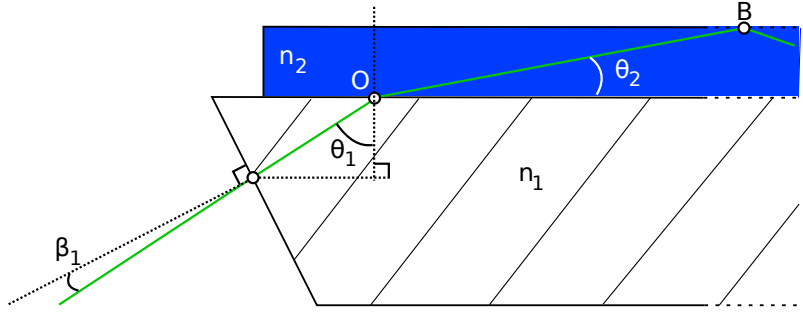}
    \caption{Scheme of the propagation of a single ray.}
    \label{fig9}
\end{figure}

All the rays are converging with respect to a spherical convex lens of focal length $f=500$ mm. Each ray independently crosses two interfaces air/PMMA and PMMA/water (Fig.\ref{fig9}). The calculation of the ray trajectory and intensity at both interface is made with respect to the Descartes Laws :

\begin{eqnarray}
\theta_{t}(\eta)=arcsin\left(\eta sin(\theta_{i})\right)\\
\label{Desc}
T(\eta)=1-R=1-{\left| \frac{tan\left(arcsin\left(\eta sin(\theta_{i})\right)-\theta_{i}\right)}{tan\left(arcsin\left(\eta sin(\theta_{i})\right)+\theta_{i}\right)}\right|}^{2}
\end{eqnarray}
where $\eta$ is the ratio between the refractive index of the incident medium and the refractive index of the transmission medium ($\eta=\frac{n^{air}}{n^{PMMA}}$ for the entry of the ray in the PMMA block and $\eta=\frac{n^{PMMA}}{n^{water}}$ at its outlet, at the interface location $O$ (Fig.\ref{fig9}).$\theta_{i}$ is the incident angle at the interface and $\theta_{t}$ is the transmitted angle. $T$ is the transmission coefficient at the interface.

The theoretical value, at $532$ nm, of the refractive index of the PMMA and the water are respectively ${n_{0}}^{PMMA}=1.49$ and ${n_{0}}^{water}=1.33$ (at $20^{\circ}C$). However, actual values of these refractive indices can vary depending on the manufacturing process for the PMMA ($n^{PMMA}={n_{0}}^{PMMA}\pm 0.01$ according to Beadie et al. 2015 \cite{beadie2015refractive}) and depending on the temperature for the water ($n^{water}={n_{0}}^{water}\pm 0.01$ according to Bashkatov and Gemina 2003 \cite{bashkatov2003water}). The reference ratio between refractive indices is expressed as $\eta_{0}=\frac{{n_{0}}^{PMMA}}{{n_{0}}^{water}}$. Thus the actual ratio can be expressed as $\eta=\eta_{0}+\delta\eta$. The variations in the transmitted angle and the transmission coefficient $T$ depend on $\eta$ as $\theta_{t}(\eta)={\theta_{t}}_{0}+\delta\theta_{t}$ and $T(\eta)=T_{0}+\delta T$ (where ${\theta_{t}}_{0}=\theta_{t}(\eta_{0})$ and $T_{0}=T(\eta_{0})$). The relative variations of both $T$ and $\theta_{t}$ can be calculated depending on $\delta\eta$ as $\frac{\delta\theta_{t}}{{\theta_{t}}_{0}}=\frac{\theta_{t}(\eta_{0}+\delta\eta)}{{\theta_{t}}_{0}}-1$ and $\frac{\delta T}{T_{0}}=\frac{T(\eta_{0}+\delta\eta)}{T_{0}}-1$.

\begin{figure}
    \centering
    \includegraphics[width=0.7\textwidth]{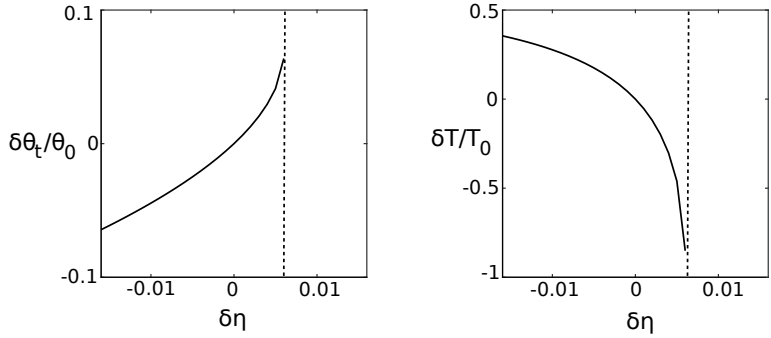}
    \caption{Relative error in the transmission angle (Left) and in the transmission coefficient (Right) as a function of the error in the refractive index ratio.}
    \label{fig10}
\end{figure}

With the uncertainty considered as $n^{PMMA}={n_{0}}^{PMMA}\pm 0.01$ and $n^{water}={n_{0}}^{water}\pm 0.01$, the variation of the ratio $\delta\eta$ lies in the range $\left[-0.016;0.016\right]$. The corresponding variations of $\frac{\delta\theta_{t}}{{\theta_{t}}_{0}}$ and $\frac{\delta T}{T_{0}}$ with $\delta\eta$ are shown in Fig.\ref{fig10}, for a single ray with an incidence $\beta_{1}=1.5^{\circ}$ on the prism.

This highlights the effect of a small variation of refractive index on a ray transmission properties. For an uncertainty of $\delta\eta=-0.016$, the variation is approximately $6\%$ for the transmitted angle $\theta_{t}$ and around $35\%$ for the transmission coefficient $T$. For $\delta\eta>0.006$ the ray is not transmitted in the water. Therefore, it is important for the experimental set-up to allow an adaptation of the incident angle to compensate a potential uncertainty on the refractive indices.

\begin{figure}
    \centering
    \includegraphics[width=0.75\textwidth]{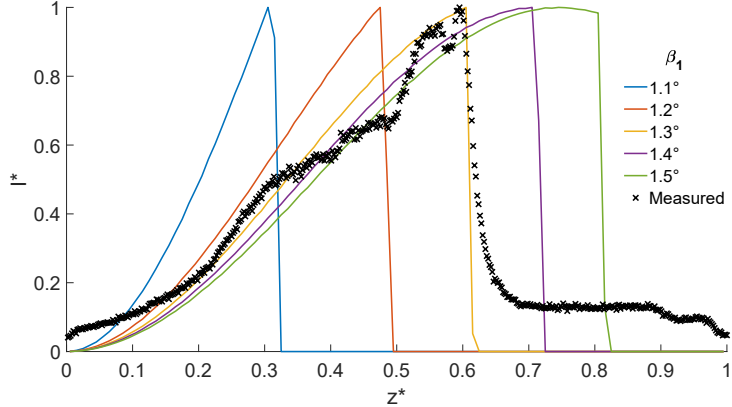}
    \caption{Normalized light intensity profile versus $\beta_{1}$ (Fig.\ref{fig9}).}
    \label{fig11}
\end{figure}

With this ray tracing model, the incident angle $\beta_{1}$ of the laser beam with the normal of the PMMA block inclined face can be modified. Therefore, the sensitivity of the intensity profile at a fixed position $x^{*}=4.455$ is investigated as a function of $\beta_{1}$ in Fig.\ref{fig11}.

\begin{figure}
    \centering
    \includegraphics[width=0.65\textwidth]{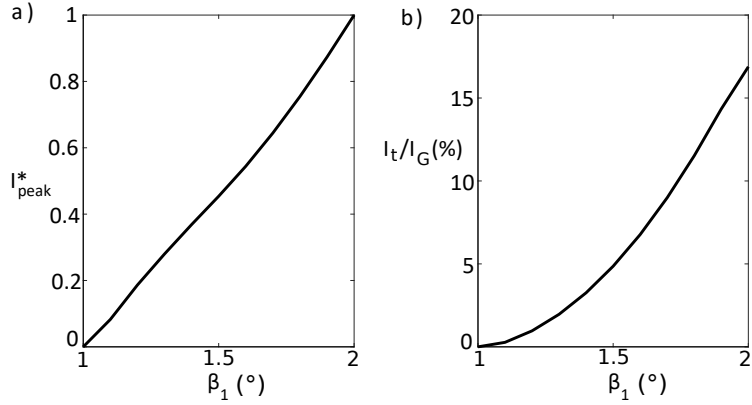}
    \caption{a) Value of the intensity peak depending on the incident angle $\beta_{1}$ (Fig.\ref{fig9}). b) Percentage of the total transmitted intensity in the fluid as a function of $\beta_{1}$.}
    \label{fig12}
\end{figure}

A good agreement of the global shape of the measured intensity profile and the intensity profile obtained with the simulation with $\beta_{1}=1.3^{\circ}$. The profiles obtained for low incident angle (around $1^{\circ}$) are thinner with a sharper peak, which is closer to the P-PIV. Therefore, the precision (thinness of the peak) of the measurement technique increases at lower incident angles. However, the intensity is here presented in its normalized form. In order to analyse the intensity of the sheet transmitted, two parameters can be analysed. The first one is the intensity peak value normalised by the intensity of the peak at $\beta_{1}=2^{\circ}$ as ${I_{peak}}^{*}=\frac{I_{peak}(\beta_{1})}{I_{peak}(\beta_{1}=2^{\circ})}$. The second is the percentage of the incident Gaussian beam transmitted in the water $I_{t}/I_{g}$ (where $I_{t}$ is the total intensity of the transmitted beam and $I_{g}$ is the total intensity of the incident Gaussian beam). These two parameters are presented Fig.\ref{fig12}.

Fig.\ref{fig12} highlights the increase of the total transmitted intensity and the intensity of the peak with the incident angle $\beta_{1}$. This ray tracing model shows us that the incident angle is a key parameter in the intensity profile structure. Considering the evolution of the thickness of the laser sheet (Fig.\ref{fig11}) with $\beta_{1}$, the incident angle should be low enough in order to increase the precision. However, considering the evolution of the intensity of the laser sheet (Fig.\ref{fig12}) with $\beta_{1}$, the incident angle should be high to ensure sufficient particles illumination and enhance the signal/noise ratio. Therefore, the chosen value for $\beta_{1}$ is a compromise between intensity and precision. This ray tracing model also highlights the high sensitivity of the method to manufacturing optical properties of the PMMA prismatic block.

\subsection{Depth of focus.}

\begin{figure}
    \centering
    \includegraphics[width=0.6\textwidth]{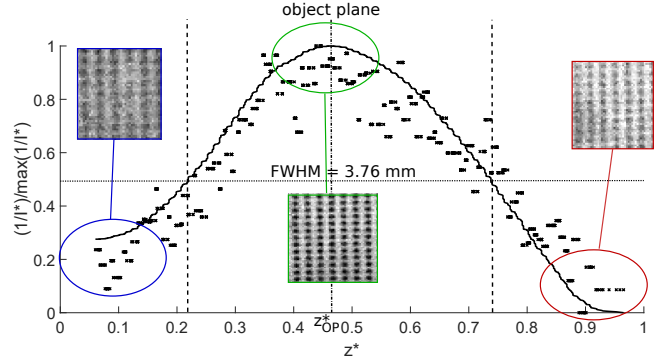}
    \caption{Normalized gray level profile for an aperture number of $5.6$.}
    \label{fig13}
\end{figure}

The optical parameter of interest for the recording optics is the depth of focus. It defines the sharpness of particle images and therefore influences the cross-correlation. To quantify this parameter, a measurement technique is developed in order to quantify the position of the object plane in the fluid at rest. A dotted target (white background plate with $0.0625$ mm diameter black dots and $0.125$ mm distance between them) is placed on an inclined plate of $45^{\circ}$ inclination with the horizontal. Images of this dotted target are recorded to determine the position $z^{*}$ of the best focused dots which are smaller and darker then the others. Therefore, the focusing quality of a dot can be calculated with their gray level. Due to the inclination of the plate the analysis of the inverse intensity of the dots ($1/I^{*}$) along the $x^{*}$ axis gives us the grey level in the $z^{*}$ direction (Fig.\ref{fig13}). This gray level is spatially averaged in the spanwise $y^{*}$ direction.

For the $5.6$ aperture number used here, the order of magnitude of the depth of focus is defined by the full width at half maximum of the gray level of the dots Fig.\ref{fig13}). This normalized depth of focus is $DOF^{*}=0.43$, which is in the same order of magnitude than the laser sheet thickness measured and shown Fig.\ref{fig7} b). With larger depth of focus, the ambient light due to the reflections of the laser sheet at the different interfaces is more visible and the signal/noise ratio is lowered. For thinner depth of focus, due to the inclination of the light sheet, the maximum intensity of the laser beam is out of the depth of focus near the images edges. Therefore, most illuminated particles are blurred and the signal/noise ratio in those parts is lowered. These are the reasons why an aperture number of $5.6$ is chosen here to ensure a good focus in the whole volume while limiting refractive noise. The object plane position (${z_{OP}}^{*}$) is then determined as the location of with the maximum gray level (Fig.\ref{fig13}), here ${z_{OP}}^{*}=0.45$.

\subsection{Conclusions.}

To conclude on optical properties of the technique, we can say that if the intensity profile illumination is initially thick, close to the impingement point at the flow window interface, the profile is thick with an asymmetrical intensity distribution. The spreading rate of the laser sheet thickness along its propagation direction is higher than for classical P-PIV for which the beam waist is usually located inside the field of view. Therefore, it appears that the local ratio between light sheet thickness and depth of focus can be a key issue for R-PIV measurements.

\section{Cross-correlation statistical model for R-PIV}
\label{CC}

\subsection{Cross correlation model (CCM).}
\subsubsection{The cross-correlation general form.}
\hfill \break

\begin{figure}
    \centering
    \includegraphics[width=0.6\textwidth]{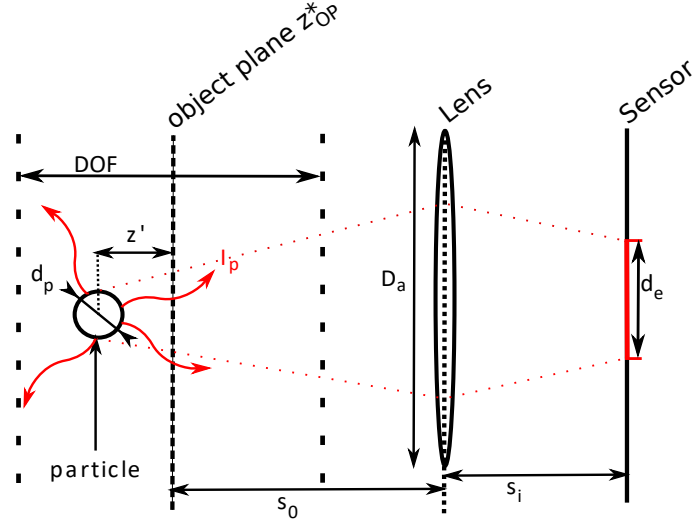}
    \caption{Scheme of the optical parameters involved to calculate a particle image diameter.}
    \label{fig14}
\end{figure}

\begin{figure}
    \centering
    \includegraphics[width=0.4\textwidth]{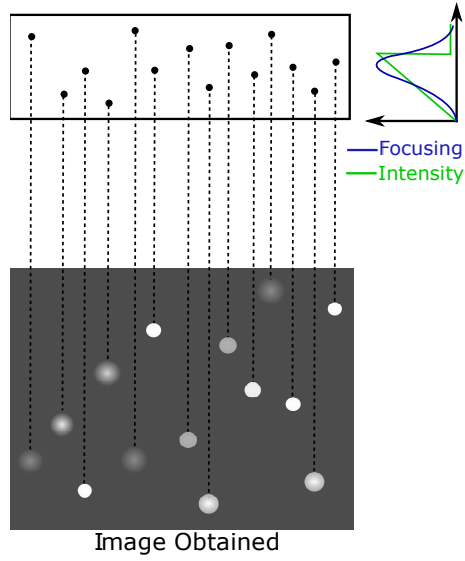}
    \caption{Scheme of the particle images depending
     on their incident intensity and focus.}
    \label{fig15}
\end{figure}

The influence of the thickness of the depth of focus (DOF) with a volume illumination has been addressed in the past revisiting the original PIV cross-correlation model of Kean and Adrian 1992 \cite{keane1992theory}. In this section, the model is revisited in order to further take into account the in-homogeneous light sheet highlighted in the previous section \ref{Profmes}. To understand the behaviour of the cross-correlation with the R-PIV optical set-up, images of individual particles are considered as a normal distribution as expressed by Olsen and Adrian 2000 \cite{olsen2000out}. Let us assume ($X$,$Y$) position variables in $x$ and $y$ directions in the sensor referential. If we consider a particle $i$ at the position (${{\mu_{1}}_{X}}_{i}$, ${{\mu_{1}}_{Y}}_{i}$) in image $1$ and (${{\mu_{2}}_{X}}_{i}={{\mu_{1}}_{X}}_{i}+{DX}_{i}$, ${{\mu_{2}}_{Y}}_{i}={{\mu_{1}}_{Y}}_{i}+{DY}_{i}$) in image 2, in the sensor referential, with ($DX_{i}$, $DY_{i}$) the displacement of this $i^{th}$ particle image on the camera sensor between both images according to Kean and Adrian 1992 \cite{keane1992theory} :

\begin{eqnarray}
{I_{1}}_{i}(X,Y)=\frac{{{I_{p}}_{1}}_{i} Da^{2}\beta^{2}}{4\pi {{{d_{e}}_{1}}_{i}}^{2}(s_{0}+{z_{1}}_{i}')^{2}}\cdot e^{-\frac{4\beta^{2}}{{{{d_{e}}_{1}}_{i}}^{2}}\left((X-{{\mu_{1}}_{X}}_{i})^{2}+(Y-{{\mu_{1}}_{Y}}_{i})^{2} \right)}\nonumber\\
{I_{2}}_{i}(X,Y)=\frac{{{I_{p}}_{2}}_{i} Da^{2}\beta^{2}}{4\pi {{{d_{e}}_{2}}_{i}}^{2}(s_{0}+{z_{2}}_{i}')^{2}}\cdot e^{-\frac{4\beta^{2}}{{{{d_{e}}_{2}}_{i}}^{2}}\left((X-{{\mu_{2}}_{X}}_{i})^{2}+(Y-{{\mu_{2}}_{Y}}_{i})^{2} \right)}
\label{int}
\end{eqnarray} 
where ${{I_{p}}_{1}}_{i}$ is the intensity emitted by the particle for image 1 (in fluorescence ${{I_{p}}_{1}}_{i}={{I_{0}}_{1}}_{i}.{d_{p}}_{i}^{2}$, with ${{I_{0}}_{1}}_{i}$ the intensity of the incident laser sheet on particle $i$ and ${d_{p}}_{i}$ is the geometrical diameter of the particle), $Da$ is the aperture diameter of the lens, $\beta^{2}$ is a constant value set to $3.67$ to best approximate the Airy diffraction (according to Olsen and Adrian 2000 \cite{olsen2000out}), $s_{0}$ is the distance between the camera and the object plane and ${z_{1}}_{i}'=|{{z_{1}}_{i}}^{*}-{z_{OP}}^{*}|$ is the distance between the particle and the object plane. ${{d_{e}}_{1}}_{i}$ is the effective particle image diameter on the sensor for image 1 calculated according to Olsen and Adrian 2000 \cite{olsen2000out} as :

\begin{eqnarray}
{{d_{e}}_{1}}_{i}=\sqrt{M^{2}{d_{p}}_{i}^{2}+5.95(M+1)^{2}\lambda_{p}^{2}{f^{\sharp}}^{2}+\frac{M^{2}{z_{1}}_{i}'^{2}{D_{a}}^{2}}{(s_{0}+{z_{1}}_{i}')^{2}}}
\label{diam}
\end{eqnarray}
where $\lambda_{p}$ is the wavelength of the light fluoresced by the particle and $f^{\sharp}$ is the aperture number of the camera lens. All those optical parameters are represented in Fig.\ref{fig14} and the particle image patterns as a function of the optical parameters are sketched in Fig.\ref{fig15}.

Eq.\ref{int} assumes that the image of a particle on the camera sensor is a 2D normal distribution. The amplitude of this Gaussian form depends on the intensity emitted by the particle which is directly linked to the illumination of the particle and on the focusing through the parameter $d_{e}$. The standard deviation of this distribution is only linked to the focusing $d_{e}$.

Considering N particles per interrogation window with background noise, the cross-correlation function displacement component can be calculated as presented in Appendix A. These calculations give us a general form :

\begin{eqnarray}
{R_{D}}_{12}(\delta_{x},\delta_{y}) &=& \sum_{i=1}^{N} \left[\frac{{{I_{0}}_{1}}_{i}.{{I_{0}}_{2}}_{i}.{d_{p}}_{i}^{4}.Da^{2}.\beta^{2}}{16\pi ({{d_{e}}_{1}}_{i}^{2}+{{d_{e}}_{2}}_{i}^{2}).(s_{0}+{z_{1}}_{i}')^{2}.(s_{0}+{z_{2}}_{i}')^{2}} \ldots \right. \nonumber \\
&\cdot & e^{-4\beta^{2}\frac{({\mu_{2}}_{X}-{\mu_{1}}_{X}-\delta_{x})^{2}+({\mu_{2}}_{Y}-{\mu_{1}}_{Y}-\delta_{y})^{2}}{{{d_{e}}_{1}}_{i}^{2}+{{d_{e}}_{2}}_{i}^{2}}}\Big]
\label{CCM_gen}
\end{eqnarray}
To reduce the equation, let's consider $A_{i}=\frac{{{d_{p}}_{i}}^{4}.Da^{2}.\beta^{2}}{16.\pi}$.

In the specific case of the channel experiment, we can assume few hypotheses to reduce the equation :

\begin{itemize}
\item Hypothesis 1 : the inclination of the laser sheet is negligible at the scale of an interrogation window. The laser sheet inclination is here about $5^{\circ}$. Therefore, the difference of maximum intensity ${z_{max}}^{*}$, between the edges of an interrogation window of size $dw_{x}^{*}=0.066$ is $\Delta {z_{max}}^{*}={z_{max}}^{*}(x^{*}+dw_{x})-{z_{max}}^{*}(x^{*})=0.005$, which is low compared to the interrogation window size $dw_{x}^{*}$. Therefore the laser sheet intensity can be considered as only dependent on $z^{*}$ in an interrogation window : ${{I_{0}}_{1}}(x^{*},y^{*},z^{*})={{I_{0}}_{1}}(z^{*})$.
\item Hypothesis 2 : both laser cavities generate perfectly aligned pulsed light sheet with the same intensity. Therefore, intensity profiles in the interrogation window for both images are equal : ${{I_{0}}_{1}}(z^{*})={{I_{0}}_{2}}(z^{*})$.
\item Hypothesis 3 : the velocity component $W$ is negligible. Therefore, the ith particle is at the same $z^{*}_{i}$ in both images. Thus, ${z_{1}}_{i}'={z_{2}}_{i}'$, ${{d_{e}}_{1}}_{i}={{d_{e}}_{2}}_{i}$ and ${{I_{0}}_{1}}(z^{*}_{i})={{I_{0}}_{2}}(z^{*}_{i})=I_{0}(z^{*}_{i})$.
\end{itemize}

With these hypotheses, the equation \ref{CCM_gen} becomes :

\begin{eqnarray}
{R_{D}}_{12}(\delta_{x},\delta_{y})&=&\sum_{i=1}^{N} \frac{A_{i}.{I_{0}}^{2}(z_{i}^{*})}{2.{d_{e}}^{2}(z_{i}^{*}).(s_{0}+z_{i}')^{4}}.e^{-4\beta^{2}\frac{({DX}(z_{i}^{*})-\delta_{x})^{2}+({DY}(z_{i}^{*})-\delta_{y})^{2}}{2.{d_{e}}^{2}(z_{i}^{*})}}\nonumber \\
&=&\sum_{i=1}^{N} {R_{D}}_{z}(z_{i}^{*})
\label{CCM_simp}
\end{eqnarray}

In conclusion, the more general form of the cross-correlation displacement component is given by the equation \ref{CCM_gen}. This equation can be simplified for the low inclination light sheet measurements for R-PIV with negligible velocity in the $z$ direction and spatial uniformity in both $x^{*}$ and $y^{*}$ directions. The simplified form (Eq\ref{CCM_simp}) is used for the channel flow studied for R-PIV testing.

Finally, to calculate the cross-correlation function for an interrogation window for a given optical set-up with the model, the inputs needed are :

\begin{itemize}
\item an intensity profile $I_{0}(z^{*})$ 
\item reception optic properties : focal length $f$, magnification $M$, working distance $s_{0}$, position of the object plane $z_{OP}^{*}$, apperture number $f^{\sharp}$ and particle reemitting wavelength $\lambda_{p}$.
\item a velocity profile ($U(z^{*})=DX(z^{*})/(M.dt)$, $V(z^{*})=DY(z^{*})/(M.dt)$)
\item a particle diameter $d_{p}$
\end{itemize}

\subsubsection{Statistical aspect and convergence.}
\hfill \break
\label{stat}

As shown by equation \ref{CCM_simp}, the cross-correlation displacement component for a single image pair is the summation of the individual cross-correlation of each particles within the interrogation volume. The cross-correlation function obtained for a single image pair is highly dependent on the random distribution of the $N$ particles in $z^{*}$ ($z^{*}_{1}$,$z^{*}_{2}$,...$z^{*}_{N}$). Therefore, this measurement technique can be highly variable for one single image pair to an other.

Therefore, to ensure the convergence of the R-PIV technique, a statistical averaging over independent image pairs should be done. In PIV, the velocity is given by finding the displacement corresponding to the location of the cross-correlation peak. For a fixed number $N_{im}$ of image pairs, an averaging strategy is used based on the correlation model equation \ref{CCM_simp}. Let's consider that the velocity finally obtained at the end of the averaging process is ($U$, $V$) with $U=\frac{\delta x^{(f)}}{M.dt}$ and $V=\frac{\delta y^{(f)}}{M.dt}$. Here ($\delta x^{(f)}$, $\delta y^{(f)}$) are the final displacements obtained after the averaging process.

This process consists in finding the location of the peak value $(\delta x,\delta y)$ of the averaged cross-correlation function over $N_{im}$ image pairs. This corresponds to the Sum of Correlation cross-correlation method.

For the averaging method, if we consider a probability law $P(z)$ of presence of a particle at a height $z$, the averaged displacement component of the cross-correlation $<{R_{D}}_{12}({\delta x},{\delta y})>$ converges towards the integral of the contributions of every $z$ when the number of particles tends to the infinity.

\begin{eqnarray}
<{R_{D}}_{12}({\delta x},{\delta y})>&=&\lim\limits_{N_{im}\rightarrow\infty} \frac{1}{N_{im}}\sum_{i=1}^{N_{im}}{R_{D}}_{12}({\delta x},{\delta y})\nonumber \\ 
&=& \int_{z_{min}}^{z_{max}} P(z) {R_{D}}_{z}(z) dz
\label{statform}
\end{eqnarray}

\begin{figure}
    \centering
    \includegraphics[width=0.9\textwidth]{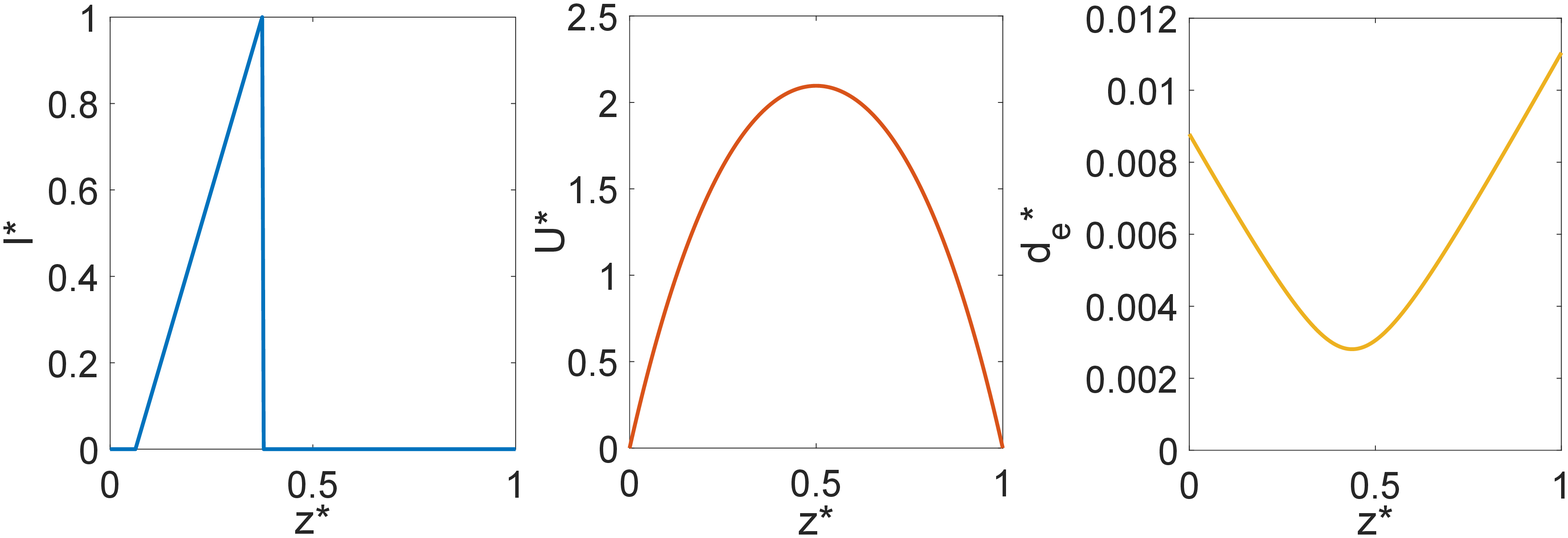}
    \caption{Inputs profiles for the CCM intensity, velocity and particle image diameter.}
    \label{fig16}
\end{figure}

To test this averaging method, an interrogation volume is considered with $N=15$ particles in this volume (approximately the number of particles per interrogation windows for a good enough velocity determination). The intensity profile considered is a simplified in-homogeneous profile to approximate the actual one measured in the channel (Fig.\ref{fig7}b)). The velocity profile is the one given by Eq.\ref{White} for $y^{*}=0.5$. The particle image diameter profile is calculated from Eq.\ref{diam} with an object plane at ${z_{OP}}^{*}=0.45$ to well simulate the experimental conditions. The input profiles are presented in Fig.\ref{fig16}.

\begin{figure}
    \centering
    \includegraphics[width=1\textwidth]{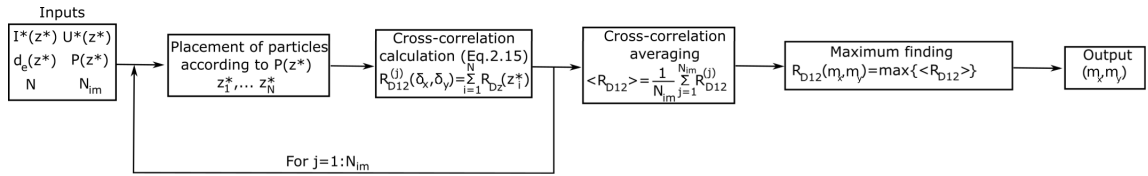}
    \caption{Scheme of the structure of the algorithm used.}
    \label{fig17}
\end{figure}

 Simulations with the model Eq.\ref{statform} are made considering an increasing number of image pairs ($N_{im}$) in order to study the convergence with $N_{im}$ of this model. The algorithm used is sketched in Fig.\ref{fig17}. To quantify this convergence, simulations are repeated $20$ times with different particle random positions for every $N_{im}$, this allows to quantify the standard deviation of the velocities given by the model depending on $N_{im}$.
 
\begin{figure}
    \centering
    \includegraphics[width=0.9\textwidth]{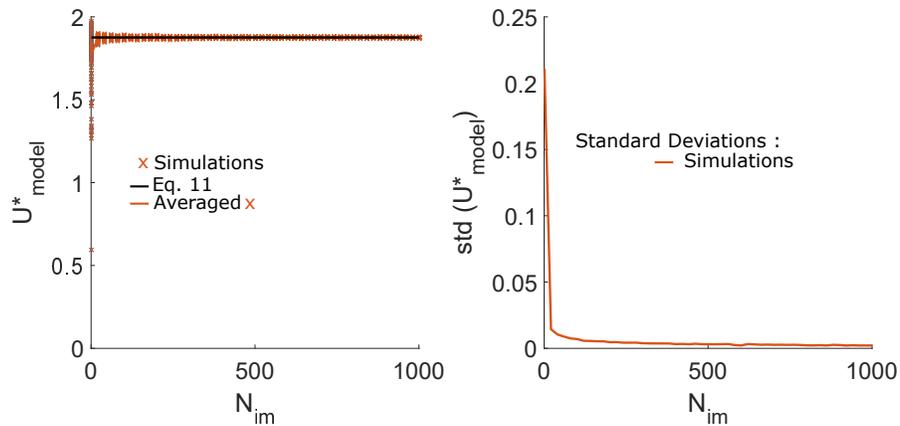}
    \caption{Simulations performed with the averaging method and its standard deviations for $N=15$.}
    \label{fig18}
\end{figure}

With this analysis, we can observe, on Fig.\ref{fig18}, that the velocity obtained with the averaging process converges to the integral value (Eq.\ref{statform}) in the limit of large $N_{im}$ values.

In conclusion, with the CCM Eq.\ref{CCM_simp}, the statistical analysis (Eq.\ref{statform}, corresponding to the Davis software (used here) sum of correlation, converges with a low dispersion (Fig.\ref{fig18} (right)). Eq.\ref{statform} can also allow the study of the converging value without repeating the simulations with the integral form. Therefore, for this study of the R-PIV, the validation of the CCM model Eq.\ref{CCM_simp} will be done comparing the measurement results using the sum of correlation averaging process with the ones predicted using the model Eq.\ref{statform}. In this article, the probability law $P(z)$ of presence of a particle is considered as homogeneous in the channel.

\subsection{Analysis of the illumination methods with the CCM}
\label{AI}
In this section the model is used in its integral form (Eq.\ref{statform}) to predict the measured velocity ($U^{*}_{model}$) in a square channel flow applying either P-PIV or R-PIV. For the P-PIV, the model is firstly used to quantify the bias while scanning the flow with two different optical configurations. The first one consists in aligning the object plane position $z_{OP}^{*}$ with the laser beam $z_{B}^{*}$ for each scanning position($z_{OP}^{*}=z_{B}^{*}$). The second one consists in scanning the flow with the laser beam $z_{B}^{*}$ with a fixed $z_{OP}^{*}=0.5$. For R-PIV, the bias introduced is quantified with a single optical configuration similar to the one described in the Section.\ref{OP}. The reconstruction of the velocity profile $U(z^{*})$ is  discussed using the inclination of the laser sheet. As described previously, the inputs needed are : the reference velocity profile, the intensity profile and the object plane location. The reference velocity field $U^{*}_{ref}$ is chosen as the analytical solution Eq.\ref{White} for $y^{*}=0.5$. For this experimental set-up, the particles are homogeneously distributed in the water tank and then the probability distribution of particles positions is supposed to be uniform in the test section. The velocity $U^{*}_{comp}$ is then compared to the reference velocity $U^{*}_{ref}$ for both illumination methods.

The relative error (${E_{1}}_{\alpha}$) is quantified to compare velocity profiles $U^{*}_{ref}$ and $U^{*}_{comp}$. This error is defined as it follows :
\begin{eqnarray}
{E_{1}}_{\alpha}=\frac{\sqrt{\int_{0}^{1}{\big(U^{*}_{ref}(\alpha)-U^{*}_{comp}(\alpha)\big)}^{2}d\alpha}}{\int_{0}^{1}U^{*}_{ref}(\alpha)d\alpha} 
\label{E1}
\end{eqnarray}
where $\alpha$ is a generic coordinate that could be either $x^{*}$, $y^{*}$ or $z^{*}$.

In this specific section, $U^{*}_{comp}=U^{*}_{model}$ and $U^{*}_{ref}$ is the velocity profile obtained with the analytical solution Eq.\ref{White} for $y^{*}=0.5$.

\subsubsection{Model Prediction for P-PIV.}
\hfill \break
\label{PPIVpred}
In the case of P-PIV, the measured horizontal laser sheet intensity profile (Fig.\ref{fig7}a)) is close to a Gaussian distribution. In this example, the thickness measured at $I_{max}^{*}/4$ is $e^{*}\simeq 0.05$.\\

\begin{itemize}
\item P-PIV scanning with $z_{OP}^{*}$ adjustment :

P-PIV is simulated using the model with a position of the object plane at $z_{OP}^{*}$ corresponding to the location of the center of the laser beam $z_{B}^{*}=z_{OP}^{*}$ (maximum intensity of the Gaussian profile). For a fixed location $x^{*}$, calculations are made with Eq.\ref{statform} with different laser beam and object plane positions from $z_{B}^{*}=z_{OP}^{*}=0.0625$ to $z_{B}^{*}=z_{OP}^{*}=0.9375$. Those simulations were made with two different light sheet thickness ($e^{*}=0.05$ and $e^{*}=0.19$), this last one in order to quantify the effect of a large light sheet. The errors between the model profiles and the analytical solution are calculated according to Eq.\ref{E1} and represented in Fig.\ref{fig19}.

\begin{figure}
    \centering
    \includegraphics[width=1\textwidth]{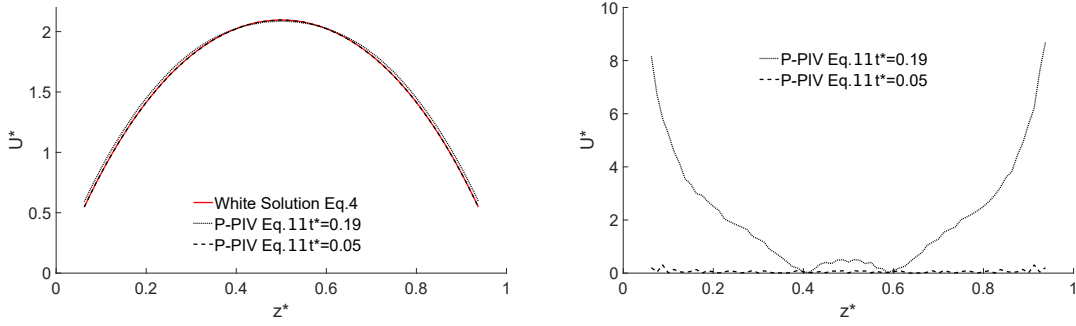}
    \caption{Left is the velocity profile $U^{*}(z^{*})$ at $y^{*}=0.5$ with the White\cite{white2006viscous} solution (Eq.\ref{White}) compared to the profile predicted by the model with different laser sheet thickness. Right is the relative error (eq.\ref{E1}) between the White\cite{white2006viscous} and the velocities predicted by the model.}
    \label{fig19}
\end{figure} 

This shows that with $z_{B}^{*}=z_{OP}^{*}$, the measurement bias introduced by the laser sheet thickness is limited. For a thick laser sheet ($e^{*}=0.19$) the error remains below $8\%$. For the thin laser sheet ($e^{*}=0.05$, as measured for the experiment), the maximum error is approximately $0.1\%$. Therefore, the P-PIV with a thin laser sheet with $z_{B}^{*}=z_{OP}^{*}$ can be considered as a reference measurement of the flow of interest.\\

\item P-PIV scanning with fixed $z_{OP}^{*}=0.5$ :

The same illumination profiles ($e^{*}=0.05$ and $e^{*}=0.19$) are studied with a fixed positioning of the object plane at $z_{OP}^{*}=0.5$, while varying the location of the light sheet from $z_{B}^{*}=0.0625$ up to $z_{B}^{*}=0.9375$. Those calculations are presented on  Fig.\ref{fig20} with the error between profiles calculated according to Eq.\ref{E1}.

\begin{figure}
    \centering
    \includegraphics[width=1\textwidth]{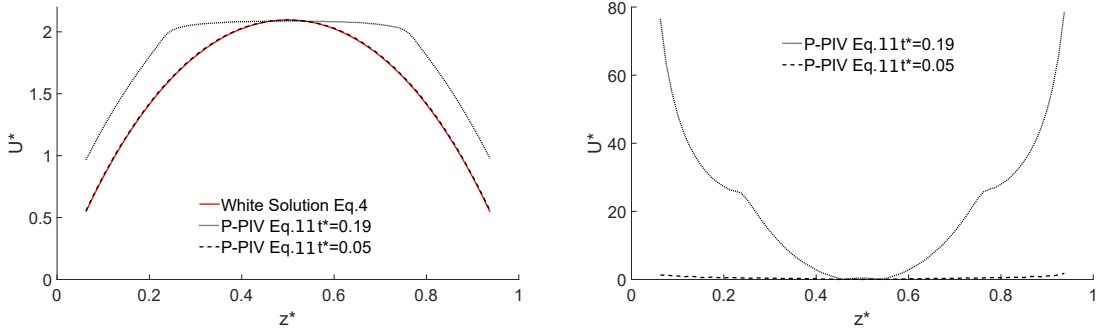}
    \caption{Left is the velocity profile $U^{*}(z^{*})$ at $y^{*}=0.5$ with the White\cite{white2006viscous} solution (Eq.\ref{White}) compared to the profile predicted by the model with different laser sheet thickness. Right is the relative error (eq.\ref{E1}) between White\cite{white2006viscous} and the velocities predicted by the model.}
    \label{fig20}
\end{figure} 

In this configuration, the error with the thin laser sheet ($e^{*}=0.05$) remains lower than $1.5\%$. However, with the thick laser sheet ($e^{*}=0.19$) the error grows up to $80\%$. This highlights the importance in P-PIV to reduce the laser sheet thickness to increase precision. When the laser sheet thickness increases, particles at different height with different velocities contribute to the cross-correlation. Therefore, the bias introduced in the measurements is increased. These preliminary results also show that the accuracy of P-PIV is increased when the object plane is aligned with the laser sheet ($z_{B}^{*}=z_{OP}^{*}$).
\end{itemize}

\subsubsection{Model Prediction for R-PIV.}
\hfill \break

\begin{figure}
    \centering
    \includegraphics[width=0.7\textwidth]{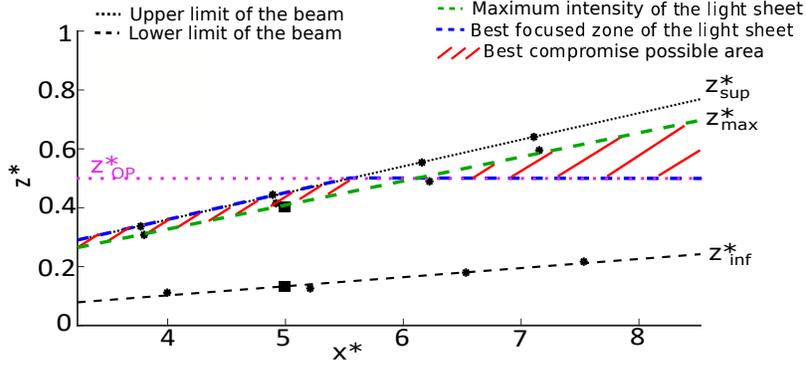}
    \caption{Propagation of the laser sheet with boundaries of the laser sheet and maximum intensity represented according to Fig..\ref{fig8} with the object plane $z^{*}_{OP}$ represented in purple. Best focused zone belonging to the light sheet is represented in blue and maximum intensity zone in green. The red area represents the zone where the best compromise between intensity $I_{0}(z^{*})$ and focusing $d_{e}(z^{*})$ is possible.}
    \label{fig21}
\end{figure}

In the case of R-PIV, the reference velocity profile $U^{*}_{ref}(z^{*})$ is fixed and invariant along $x^{*}$. The particle image diameter profile $d_{e}(z_{i}^{*})$ is also invariant along $x^{*}$ due to the fixed object plane position at $z_{OP}^{*}=0.5$. The only input of the model Eq.\ref{statform} which varies along $x^{*}$ is here the intensity profile $I_{0}(z_{i}^{*})$. In Eq.\ref{CCM_simp}, we can observe that the amplitude the final Gaussian depends on each $I_{0}(z_{i}^{*})$ squared and $d_{e}(z_{i}^{*})$ squared corresponding to the contribution of the $i^{th}$ particle. As shown on Fig.\ref{fig8} when $z^{*}_{max}$ is below $z^{*}_{OP}$, the best focused particles (close to $z_{OP}^{*}$) are the ones with the highest intensities. Thus, we can assume here that the velocity given by the R-PIV is the one at the position of the maximum intensity $z^{*}_{max}$ on the vertical. In the zone where $z^{*}_{max}>z_{OP}^{*}$, it is difficult to determine which height provide the higher ratio between intensity and focusing (image diameter). Therefore, it is difficult to associate the velocity given by the R-PIV to a specific location $z^{*}$. This is illustrated by the red dashed area in Fig.\ref{fig21}. When $z^{*}_{max}<z_{OP}^{*}$, the potential best compromise area (higher ratio) is thin and close to $z^{*}_{max}$. The thickness of this area increases when $z^{*}_{max}>z_{OP}^{*}$.

The cross-correlation model calculation can be repeated for every $x^{*}$ from $0$ to $6.3$ (location for which $z^{*}_{max}(x^{*})=z_{OP}^{*}$). The velocity given by the model $U^{*}_{model}$ is then associated to the maximum intensity location $z^{*}_{max}(x^{*})$. Thus the model velocity profile $U^{*}_{model}(z^{*}_{max}(x^{*}))$ is obtained up to $z^{*}_{max}(x^{*})=0.5$. Then, knowing the symmetry reference velocity profile, this property is used to extrapolate the velocity profile $U^{*}_{model}$ for $z^{*}>0.5$.

\begin{figure}
    \centering
    \includegraphics[width=1\textwidth]{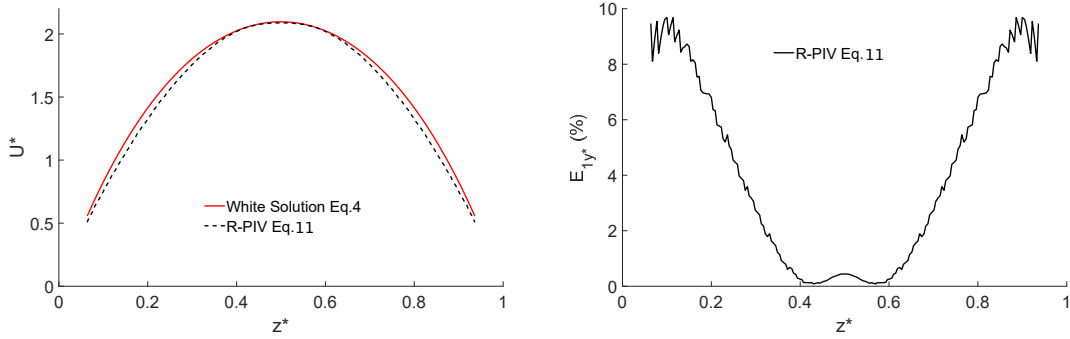}
    \caption{Left is the velocity profile $U^{*}(z^{*})$ at $y^{*}=0.5$ with the White and Corfield 2006 \cite{white2006viscous} solution (Eq.\ref{White})compared to the profile predicted by the model with refracted laser beam. Right is the error (eq.\ref{E1}) between White and Corfield 2006 \cite{white2006viscous} at $z^{*}_{max}$ and the velocity predicted by the correlation model for R-PIV.}
    \label{fig22}
\end{figure}

The reference velocity profile $U^{*}_{ref}$ is compared to $U^{*}_{model}$ in Fig.\ref{fig22}. The relative error is given by Eq.\ref{E1} with $U^{*}_{ref}=U^{*}_{ref}(z^{*}_{max})$ and $U^{*}_{model}=U^{*}_{comp}$.

With a refracted light sheet and a fixed object plane at $z_{OP}^{*}=0.5$, the error between the profiles obtained by the model and the reference profiles (analytical solution) is below $1$ \% in the low velocity gradient ($\frac{dU^{*}}{dz^{*}}$) part of the flow. However, the error grows rapidly near the walls with an error of approximately $9$\% at $z^{*}=0.0625$. 
It is interesting to note that this bias is due to the velocity gradient that is accounted in the model. Thus R-PIV induces an intrinsic (due to optical-correlation process of analysis) error that is not negligible, contrarily to P-PIV, as soon as the studied flow shows large gradient.

Now, the model can be used to analyse the R-PIV measurements when used to characterize all the flow.

\section{R-PIV measurement results}
\label{MR}
\subsection{Flow Structure using P-PIV.}

Due to the geometry constraints of the experimental set-up shown in Fig.\ref{fig4}, the measurement area is centred around $148$ mm downstream of the channel inlet. The ratio $L/h$ between the length $L$ of the channel and its height $h$ is around $18.5$. This distance is too short to consider a fully developed channel flow. In these conditions, P-PIV measurements can locally show velocities which are not fully in accordance with the analytic solution (Eq.\ref{White}). The three main reasons that can explain these differences are : 1) the bias introduced by the laser sheet thickness (negligible according to Section \ref{PPIVpred}), 2) the disturbance of the flow due to the set-up imperfections as the adjustment of the connector at the inlet of the channel and 3) the not fully developed flow 4) local disturbances of the flow at the exit of the test section.

Here for the P-PIV measurements, the properties of the laser light sheet used is as presented on Fig.\ref{fig7}a) with a thickness of $e^{*}=0.05$ and with an object plane parallel to the laser sheet ($z_{OP}^{*}=z_{B}^{*}$). As shown previously, in Section \ref{PPIVpred} for these conditions, the bias introduced by the laser sheet thickness is negligible. Therefore, the small differences between P-PIV measurements and the analytical solution for such flow are due mainly to the flow disturbances.

\begin{figure}
    \centering
    \includegraphics[width=0.7\textwidth]{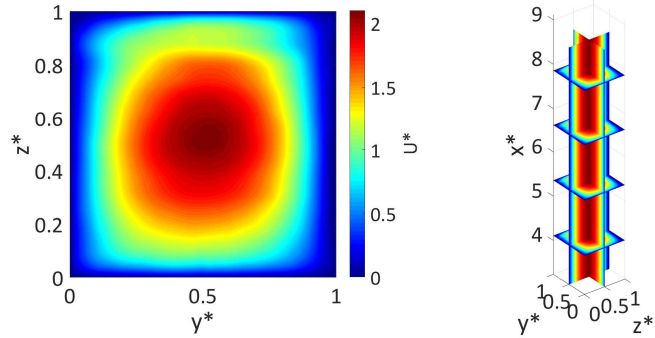}
    \caption{Measured velocity field in the channel (cubic spline interpolation from 2D slices obtained with P-PIV).}
    \label{fig23}
\end{figure}

P-PIV is here used first to characterize the flow field in the channel. This measurement method is considered as a reference as its robustness and limitations were extensively demonstrated in the literature for similar flows. To obtain the most complete velocity field representation of the channel flow, P-PIV measurements are done by scanning the position $z^{*}_{B}$ of the laser sheet with a step of $0.5$ mm from $z_{B}^{*}=0.0625$ to $z_{B}^{*}=0.9375$. The object plane location is adjusted for every $z^{*}_{B}$ in order to ensure that $z_{OP}^{*}=z_{B}^{*}$. A spatial interpolation is finally made to reconstruct the 3D2C velocity field ($U^{*}(x^{*},y^{*},z^{*})$) in the channel (Fig.\ref{fig23}) (interpolation with cubic spline method). Measurements in a cross-section show some local departure from a perfectly symetric flow with respect to ($x^{*}$,$y^{*}$) and ($x^{*}$,$z^{*}$) mid-planes. The streamwise inhomogeneity appears to be weak.

\subsection{R-PIV results.}
\label{R-PIV}

With a refracted light sheet illumination as presented in Fig.\ref{fig8}, the PIV analysis is leaded with a sum of correlation processing over $500$ recorded image pairs in order to converge to the integral formulation (Eq.\ref{statform}). The velocity is then determined by finding the peak location of this averaged correlation. The object plane for these measurements is fixed at $z_{OP}^{*}=0.45$ (Fig.\ref{fig13}). 

\begin{figure}
    \centering
    \includegraphics[width=1\textwidth]{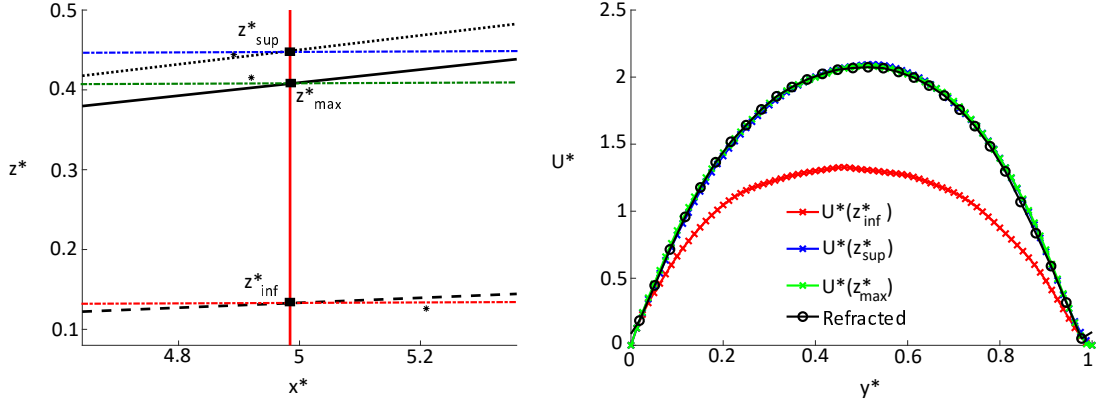}
    \caption{Left : Zoom on the section at $x^{*}=5$ from Fig.\ref{fig8}. Right : Colored lines with cross markers are profiles obtained with P-PIV at fixed $z^{*}$, these colors correspond to the colors for the various $z^{*}$ horizontal sheets in the left figure. The black line with circle markers represents the profile measured with R-PIV.}
    \label{fig24}
\end{figure}

The velocity profile in the $y^{*}$ direction, obtained with the R-PIV, is compared to the P-PIV reference profiles at the specific heights $z_{inf}^{*}$, $z_{max}^{*}$ and $z_{sup}^{*}$ (Fig.\ref{fig8}). This comparison was made for every section $x^{*}$ in the measurement area. An example is presented in Fig.\ref{fig24} for $x^{*}=5$ already highlighted Fig.\ref{fig8}.

\begin{figure}
    \centering
    \includegraphics[width=0.75\textwidth]{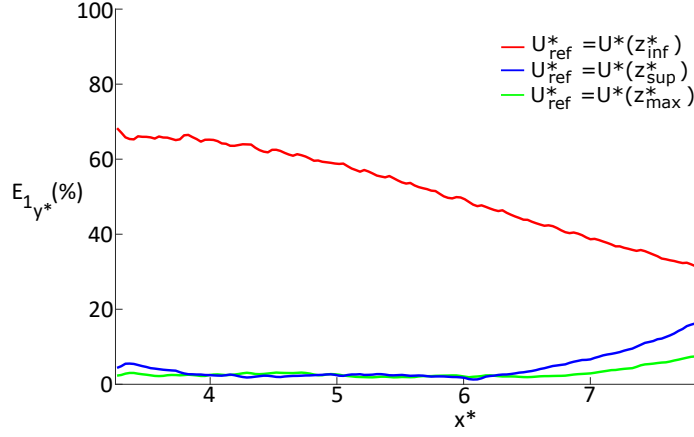}
    \caption{The error Eq.\ref{E1} is represented with $U^{*}_{comp}=U_{R-PIV}^{*}$ and $U^{*}_{ref}=U^{*}_{P-PIV}(z^{*}_{inf})$ (Red), $U^{*}_{ref}=U^{*}_{P-PIV}(z^{*}_{sup})$ (Blue) and $U^{*}_{ref}=U^{*}_{P-PIV}(z^{*}_{max})$  (Green).}
    \label{fig25}
\end{figure}

Such comparisons demonstrate that the velocities measured with R-PIV are in good agreement with the P-PIV horizontal measurements at the height $z_{max}^{*}$. This demonstrates that with an inhomogeneous and asymmetrical light sheet, the particles which are the most illuminated contribute more to the cross-correlation. To quantify the validity of this conclusion over the whole measurement area, the global error indicator $E_{1y^{*}}$ between the reference profiles ($U^{*}_{ref}=U^{*}_{P-PIV}(z_{inf}^{*})$, $U^{*}_{ref}=U^{*}_{P-PIV}(z_{max}^{*})$ and $U^{*}_{ref}=U^{*}_{P-PIV}(z_{sup}^{*})$) and the R-PIV measurements ($U^{*}_{comp}=U^{*}_{R-PIV}$) is calculated according to the Eq.\ref{E1} at every section $x^{*}$ (Fig.\ref{fig25}).

Fig.\ref{fig25} confirms that the R-PIV measurements are close to the velocity obtained with P-PIV at $z_{max}^{*}$. The Fig.\ref{fig25} (green curve) shows that for $x^{*}\geqslant 6$ the error between the R-PIV velocity measurements and the P-PIV velocity measurements at $z_{max}^{*}$ grows. Remember that for every $x^{*}$ is associated a unique $z_{max}^{*}$. For $x^{*}\geqslant 6$, the particles at $z_{max}^{*}$ are not the closest of the object plane. Some particles at lower altitude are illuminated due to the asymmetry of the intensity profile and are best focused. Therefore the contribution of these particles is more important than the ones at $z_{max}^{*}$ and the velocity field measured will correspond to the velocity at the best focused height.

\begin{figure}
    \centering
    \includegraphics[width=0.9\textwidth]{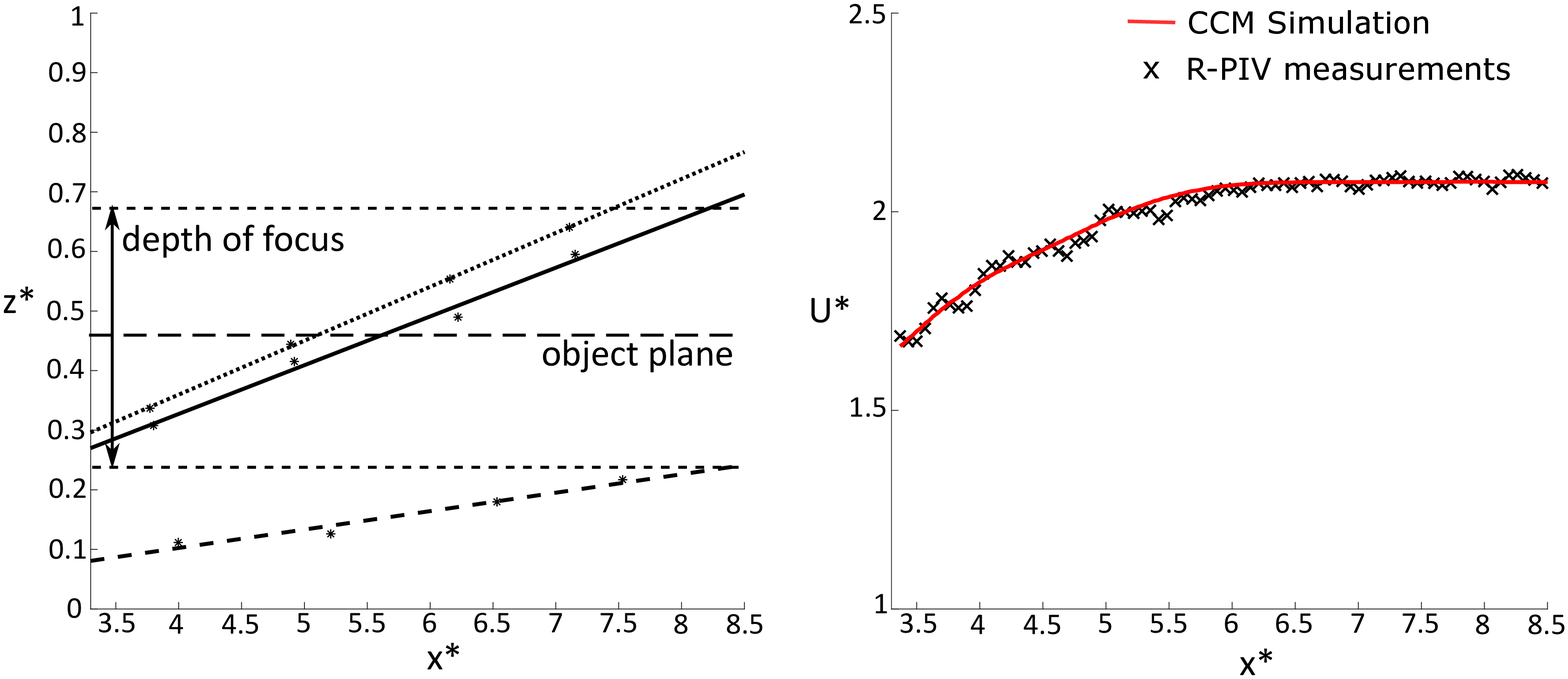}
    \caption{Left side is the laser sheet evolution in $x^{*}$ with the object plane ($z_{OP}^{*}$) and the depth of focus shown. Right side is the velocity at $y^{*}=0.5$ for the R-PIV measurement (black points) and the velocity obtained with the model for the same position (red line).}
    \label{fig26}
\end{figure}

We have further ran the model Eq.\ref{statform} with the averaging integral formulation considering the actual experimental parameters (Intensity profile, object plane position and reference velocity profile deduced from P-PIV scanning). This predicted velocity is compared in Fig.\ref{fig26} to the measured one using R-PIV (with the sum of correlation over the 500 image pairs) for various $x^{*}$ at a fixed $y^{*}=0.5$.

This shows a good agreement between the R-PIV measurements and the predicted velocity by the model. Therefore, the optical model Eq.\ref{CCM_simp} can be used with the averaging process Eq.\ref{statform} to describe the bias in the R-PIV measurements depending on the optical parameters of the experiment.

\begin{figure}
    \centering
    \includegraphics[width=0.8\textwidth]{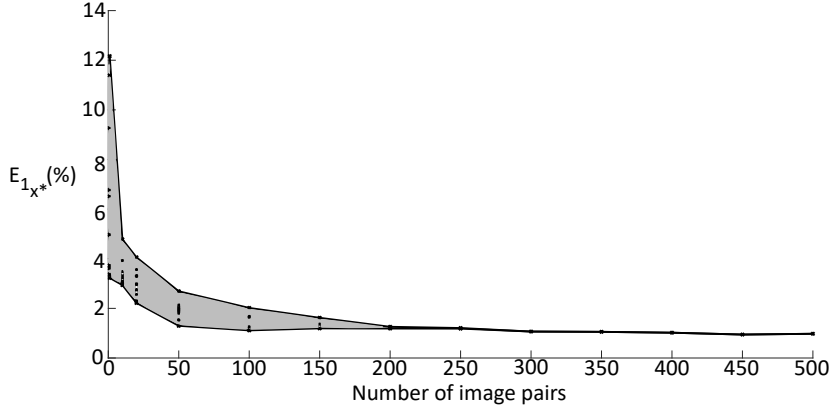}
    \caption{Error eq.\ref{E1} is represented with $U^{*}_{ref}$ the velocity measured with R-PIV and $U^{*}_{comp}$ the velocity predicted by the model.}
    \label{fig27}
\end{figure}

As discussed in Section \ref{statform}, the averaging process Eq.\ref{statform} is a statistical interpretation of the sum of correlation when the number of image pairs considered tends to infinity. To check this statistical convergence with measurements, the R-PIV image pairs are cross-correlated with the exact same process considering a number of image pairs growing from 1 to 500. To quantify the error between the R-PIV measured velocity with the prediction of the CCM, the relative error is calculated according to Eq.\ref{E1} using $U_{ref}^{*}=U_{model}^{*}$ (red line Fig.\ref{fig26}) and $U_{comp}^{*}=U^{*}_{R-PIV}$ (black crosses Fig.\ref{fig26}). This error is represented in Fig.\ref{fig27} as a function of the number of image pairs.

This shows the convergence of $U^{*}_{R-PIV}$ towards $U_{model}^{*}$ when $N_{im}$ increases. To obtain measurements that can be interpreted and predicted by the model, the number of image pairs to consider for measurements should be over 200 image pairs.

\section{Conclusions}

\subsection{Optical properties of R-PIV}

In this paper, the R-PIV technique based on the refraction of the laser sheet at the flow window interface has been presented in order to measure velocity fields in confined flows at macroscopic scale when only one optical access is available.

With this technique, the optical properties are specific and should be measured in order to determine the accuracy of measurements. 

A measurement technique of the laser light sheet profile in the water flow was proposed in addition to a ray tracing model in order to characterize the light sheet propagation. After the refraction of the laser sheet at the solid/fluid interface, an inclined laser sheet emerges with a low inclination. This light sheet is characterized by a large thickness growing with the propagation of the laser sheet along the direction $x^{*}$ and by an asymmetrical form that remains unchanged while it grows.

\subsection{Cross-Correlation model}

This specific illumination together with the positioning of the object plane of the camera influences the shape of the cross-correlation during PIV analysis. An extension of the cross-correlation models proposed in the literature, for $\mu$-PIV, and accounting for large depths of focus is developed here for macroscopic scale measurements with an inhomogeneous incident light sheet intensity profile. The proposed model helps us to understand the integration effect intrinsic of the present R-PIV technique. This CCM model allows to study the PIV cross-correlation with an intensity profile $I_{0}^{*}(z^{*})$ to be determined experimentally for the specific optical configuration. The validity of the model has been established in simplified academic flow. This model also allows to study the convergence while using an ensemble averaging process as the sum of correlation.

The use of the R-PIV technique is then appropriated to measure the velocity field as for the tire rolling in a puddle (Cabut et al. 2019 \cite{cabut2019particle}), thin film moving on angular plates,etc. The cross-correlation model can for such extended PIV method to determine effect leading to errors.

For the specific tire application, a limitation can be the dimension of the field of view (approximately $200$ mm) for which multiple reflections of the light sheet at the free surface and on the floor can occur and modify the structure of the laser sheet. Thus, in this case, the light intensity profile should be measured at every position in the puddle to determine the exact intensity profile after the multiple reflections.

\section*{Acknowledgments :}
	
The authors would like to thank BPI France (grant n$^{\circ}$ DOS0051329/00) and R\'egion Auvergne-Rh\^one-Alpes (grant n$^{\circ}$ 16 015011 01) for funding the Hydrosafe Tire FUI project.

\clearpage
\newpage

\section*{Appendix A :}
\label{cross_corr}

Let us consider images as a summation of N particle images with a background noise as :

\begin{eqnarray}
  \left\{
    \begin{array}{l}
		I_{1}(X,Y)=\sum_{i=1}^{N}{I_{1}}_{i}(X,Y) + B_{1}\\
		I_{2}(X,Y)=\sum_{i=1}^{N}{I_{2}}_{i}(X,Y) + B_{2}
    \end{array}
  \right.  
  \label{Isom}
\end{eqnarray}
where $B1$ and $B2$ background noises respectively in image 1 and 2. ${I_{1}}_{i}$ and ${I_{2}}_{i}$, images of the particle i in image 1 and 2 according to Eq.\ref{int}.

The cross-correlation function can be calculated according to Eq.\ref{eq21} with $X=X_{w}+x$ and $Y=Y_{w}+y$ :

\begin{eqnarray}
R_{12}(\delta x,\delta y)&=&\int_{-\Delta w_{X}/2}^{\Delta w_{X}/2}\int_{-\Delta w_{Y}/2}^{\Delta w_{Y}/2}\left(\sum_{i=1}^{N}{I_{1}}_{i}(X, Y)+B_{1}\right) \nonumber \\
&\cdot & \left(\sum_{i=1}^{N}{I_{2}}_{i}(X+\delta x,Y+\delta y)+B_{2}\right) dXdY
\end{eqnarray}

Thus, the cross-correlation decomposition of Adrian 1988 \cite{adrian1988double} appears as :

\begin{eqnarray}
R_{12}(\delta x,\delta y)=\overbrace{\int_{-\Delta w_{X}/2}^{\Delta w_{X}/2}\int_{-\Delta w_{Y}/2}^{\Delta w_{Y}/2}\sum_{i=j=1}^{N}{I_{1}}_{i}(X, Y){I_{2}}_{j}(X+\delta x,Y+\delta y)dXdY}^{{R_{D}}_{12}}\nonumber \\
+\overbrace{\int_{-\Delta w_{X}/2}^{\Delta w_{X}/2}\int_{-\Delta w_{Y}/2}^{\Delta w_{Y}/2}\sum_{i\neq j}^{N}{I_{1}}_{i}(X, Y){I_{2}}_{j}(X+\delta x,Y+\delta y)dXdY}^{{R_{F}}_{12}} \\
+\overbrace{\int_{-\Delta w_{X}/2}^{\Delta w_{X}/2}\int_{-\Delta w_{Y}/2}^{\Delta w_{Y}/2} \left(B_{1}.\sum_{i=1}^{N}{I_{2}}_{i}(X+\delta x,Y+\delta y)+\sum_{i=1}^{N}{I_{1}}_{i}(X, Y).B_{2}+B_{1}.B_{2}\right)}^{{R_{C}}_{12}} dXdY \nonumber
\end{eqnarray}

Let us assume that the displacement component ${R_{D}}_{12}$ is of an higher magnitude than the two others. Thus the following cross-correlation analysis is focused on the ${R_{D}}_{12}$ component.

\begin{eqnarray}
{R_{D}}_{12}=\sum_{i=j=1}^{N}\int_{-\Delta w_{X}/2}^{\Delta w_{X}/2}\int_{-\Delta w_{Y}/2}^{\Delta w_{Y}/2}{I_{1}}_{i}(X, Y){I_{2}}_{j}(X+\delta x,Y+\delta y)dXdY \nonumber
\end{eqnarray}

With ($\gamma=X+\delta x$, $\zeta=Y+\delta y$), this cross correlation becomes :

\begin{eqnarray}
{R_{D}}_{12}=\sum_{i=j=1}^{N}\int_{-\Delta w_{X}/2+\delta x}^{\Delta w_{X}/2+\delta x}\int_{-\Delta w_{Y}/2+\delta y}^{\Delta w_{Y}/2+\delta y}{I_{1}}_{i}(\gamma-\delta x, \zeta-\delta y){I_{2}}_{j}(\gamma, \zeta)d\gamma d\zeta \nonumber
\end{eqnarray}

Taking ${I_{3}}_{i}(\gamma,\zeta)={I_{1}}_{i}(-\gamma,-\zeta)$, the displacement component becomes :

\begin{eqnarray}
{R_{D}}_{12}&=&\sum_{i=j=1}^{N}\int_{-\Delta w_{X}/2+\delta x}^{\Delta w_{X}/2+\delta x}\int_{-\Delta w_{Y}/2+\delta y}^{\Delta w_{Y}/2+\delta y}{I_{3}}_{i}(\delta x-\gamma, \delta y-\zeta){I_{2}}_{j}(\gamma, \zeta)d\gamma dv \nonumber \\
&=&\sum_{i=j=1}^{N}\left[{I_{3}}_{i}*{I_{2}}_{i}\right]\nonumber \\
&=&\sum_{i=j=1}^{N} \mathcal{F}^{-1}(\mathcal{F}({I_{3}}_{i}(\delta x, \delta y))\cdot \mathcal{F}({I_{2}}_{i}(\delta x, \delta y)))
\end{eqnarray}
where $\mathcal{F}$ is the Fourier transform symbol.

${I_{2}}_{i}$ and ${I_{3}}_{i}$ can be calculated according to Eq.\ref{int} as normal distributions with an amplitude term $C$ and a standard deviation $\sigma$ defined for both images respectively as :

\begin{eqnarray}
  \left\{
    \begin{array}{l}
      {C_{1}}_{i}=\frac{{{I_{p}}_{1}}_{i} Da^{2}\beta}{4\sqrt{\pi} {{{d_{e}}_{1}}_{i}}(s_{0}+z_{1}')^{2}}\\
      {C_{2}}_{i}=\frac{{{I_{p}}_{2}}_{i} Da^{2}\beta}{4\sqrt{\pi} {{{d_{e}}_{2}}_{i}}(s_{0}+z_{2}')^{2}}
    \end{array}
  \right.
  \Rightarrow
  \left\{
    \begin{array}{l}
      {\sigma_{1}}_{i}=\frac{{{d_{e}}_{1}}_{i}}{\sqrt{8}\beta}\\
      {\sigma_{2}}_{i}=\frac{{{d_{e}}_{2}}_{i}}{\sqrt{8}\beta}
    \end{array}\nonumber
  \right.  
  \label{C1}
\end{eqnarray}

Thus the intensity ${I_{2}}_{i}$ and ${I_{3}}_{i}$ are calculated as :

\begin{eqnarray}
  \left\{
    \begin{array}{l}
      {I_{2}}_{i}(X,Y)=\frac{{C_{2}}_{i}}{\sqrt{2\pi}{\sigma_{2}}_{i}}\cdot e^{-\left((X-{\mu_{2}}_{X})^{2}+(Y-{\mu_{2}}_{Y})^{2} \right)/(2{\sigma_{2}}_{i}^{2})}\nonumber \\
      {I_{3}}_{i}(X,Y)={I_{1}}_{i}(-X,-Y)=\frac{{C_{1}}_{i}}{\sqrt{2\pi}{\sigma_{1}}_{i}}\cdot e^{-\left((-X-{\mu_{1}}_{X})^{2}+(-Y-{\mu_{1}}_{Y})^{2} \right)/(2{\sigma_{1}}_{i}^{2})}
    \end{array}
  \right.
\end{eqnarray}

After Fourier transform calculations, the cross-correlation displacement component becomes :

\begin{eqnarray}
R_{12}(\delta x,\delta y) = \sum_{i=1}^{N} {C_{1}}_{i} {C_{2}}_{i}\cdot \frac{{\sigma_{1}}_{i}{\sigma_{2}}_{i}}{{\sigma_{1}}_{i}^{2}+{\sigma_{2}}_{i}^{2}} \cdot e^{-\left(({\mu_{2}}_{X}-{\mu_{1}}_{X}-\delta_{x})^{2}+({\mu_{2}}_{Y}-{\mu_{1}}_{Y}-\delta_{y})^{2}\right)/(2({\sigma_{1}}_{i}^{2}+{\sigma_{2}}_{i}^{2}))}\nonumber
\label{R12}
\end{eqnarray}

Substituting ${C_{1}}_{i}$, ${C_{2}}_{i}$, ${\sigma_{1}}_{i}$ and ${\sigma_{2}}_{i}$ to Eq.\ref{R12}, we finally obtain :

\begin{eqnarray}
{R_{D}}_{12}(\delta_{x},\delta_{y}) &=& \sum_{i=1}^{N} \left[\frac{{{I_{p}}_{1}}_{i}.{{I_{p}}_{2}}_{i}.Da^{2}.\beta^{2}}{16\pi ({{d_{e}}_{1}}_{i}^{2}+{{d_{e}}_{2}}_{i}^{2}).(s_{0}+{z_{1}}_{i}')^{2}.(s_{0}+{z_{2}}_{i}')^{2}} \ldots \right. \nonumber \\
&\cdot & e^{-4\beta^{2}\frac{({\mu_{2}}_{X}-{\mu_{1}}_{X}-\delta_{x})^{2}+({\mu_{2}}_{Y}-{\mu_{1}}_{Y}-\delta_{y})^{2}}{{{d_{e}}_{1}}_{i}^{2}+{{d_{e}}_{2}}_{i}^{2}}}\Big]
\end{eqnarray}

Thus the Eq.\ref{CCM_gen} have been demonstrated in this appendix.

\end{document}